\documentclass[iop]{emulateapj}
\slugcomment{{\sc Accepted to ApJ:} October 14, 2014}
\usepackage{graphicx}
\usepackage{natbib}
\usepackage{latexsym}
\usepackage{amssymb}
\usepackage{longtable}
\usepackage{amsmath}
\usepackage{url}
\def\hei{He I $\lambda$10830 }
\def\msun{\ifmmode {\rm\,M_\odot}\else ${\rm\,M_\odot}$\fi}
\def\Msun{\ifmmode {\rm\,\it{M_\odot}}\else ${\rm\,M_\odot}$\fi}
\def\lsun{\ifmmode {\rm\,L_\odot}\else ${\rm\,L_\odot}$\fi}
\def\Lsun{\ifmmode {\rm\,\it{L_\odot}}\else ${\rm\,L_\odot}$\fi}
\def\rsun{\ifmmode {\rm\,R_\odot}\else ${\rm\,R_\odot}$\fi}
\def\Rsun{\ifmmode {\rm\,\it{R_\odot}}\else ${\rm\,R_\odot}$\fi}
\def\Tsun{\ifmmode {\rm\,T_\odot}\else ${\rm\,T_\odot}$\fi}
\def\arcsec{\ifmmode {^{\prime\prime}}\else $^{\prime\prime}$\fi}
\def\asec{\ifmmode {^{\prime\prime}}\else $^{\prime\prime}$\fi}
\def\arcmin{\ifmmode {^{\prime}}\else $^{\prime}$\fi}
\def\amin{\ifmmode {^{\prime}}\else $^{\prime}$\fi}
\def\simlt{\mathrel{\spose{\lower 3pt\hbox{$\mathchar"218$}}
     \raise 2.0pt\hbox{$\mathchar"13C$}}}
\def\simgt{\mathrel{\spose{\lower 3pt\hbox{$\mathchar"218$}}
\     \raise 2.0pt\hbox{$\mathchar"13E$}}}
\def\ltsima{$\; \buildrel < \over \sim \;$}

\usepackage[pdftex,backref,breaklinks,colorlinks,citecolor=blue]{hyperref}
\usepackage[all]{hypcap}

\begin{document}

\title{Diagnosing Mass Flows Around Herbig Ae/Be Stars Using the He I $\lambda$10830 Line}
\author{P. Wilson Cauley}
\email{pcauley@wesleyan.edu}
\affil{Wesleyan University}
\affil{Department of Astronomy, 45 Wyllys Avenue, Middletown, CT 06459}
\author{Christopher M. Johns--Krull}
\email{cmj@rice.edu}
\affil{Rice University}
\affil{Department of Physics and Astronomy, 6100 Main St., MS 108, Houston, TX 77005}

\begin{abstract} 

We examine \hei profile morphologies for a sample of 56 Herbig Ae/Be stars (HAEBES). We find
significant differences between HAEBES and CTTSs in the statistics of both blue--shifted absorption
(i.e. mass outflows) and red--shifted absorption features (i.e. mass infall or accretion). Our
results suggest that, in general, Herbig Be (HBe) stars do not accrete material from their inner
disks in the same manner as CTTSs, which are believed to accrete material via magnetospheric
accretion, while Herbig Ae (HAe) stars generally show evidence for magnetospheric accretion. We find
no evidence in our sample of narrow blue--shifted absorption features which are typical indicators
of inner disk winds and are common in \hei profiles of CTTSs. The lack of inner disk wind signatures
in HAEBES, combined with the paucity of detected magnetic fields on these objects, suggests that
accretion through large magnetospheres which truncate the disk several stellar radii above the
surface is not as common for HAe and late--type HBe stars as it is for CTTSs. Instead, evidence is
found for smaller magnetospheres in the maximum red--shifted absorption velocities in our HAEBE
sample. These velocities are, on average, a smaller fraction of the system escape velocity than is
found for CTTSs, suggesting accretion is taking place closer to the star. Smaller magnetospheres,
and evidence for boundary layer accretion in HBe stars, may explain the less common occurrence of
red--shifted absorption in HAEBES. Evidence is found that smaller magnetospheres may be less
efficient at driving outflows compared to CTTS magnetospheres.

\end{abstract}


\section{INTRODUCTION}

One of the most important aspects of pre--main sequence (PMS) stellar evolution is the interaction
of the star with its surrounding environment. The rate at which the central object
accretes and ejects material, stellar and circumstellar magnetic fields, and the geometry of the
circumstellar environment all play an important role in the system's evolution and the ultimate
formation of planets. While many important details of a star's interaction with its environment
have been clarified for the low--mass PMS classical T--Tauri stars (CTTSs) \citep[e.g.
see][]{bouvier07}, much of the picture is still uncertain for the intermediate--mass PMS
Herbig Ae/Be stars (hereafter HAEBES).   

HAEBES are intermediate mass (2--10 \Msun) stars first classified by George Herbig in 1960 according
to their A-- or B--type emission line spectra and their association with, and illumination of,
nebulosity \citep{herbig60}. These criteria have since been adjusted by various authors in order to
incorporate potential HAEBES that may have been excluded based on Herbig's original criteria
\citep[e.g.][]{fink84,the94,malfait98}. A decade after Herbig's catalog was established, \citet{strom72}
confirmed the pre--main sequence nature of HAEBES by showing that their surface gravities are lower
than main sequence stars of the same spectral types. These observational results placed the HAEBES
in the HR diagram above the zero--age main sequence of theoretical evolutionary tracks \citep[e.g.
those of][]{iben}, further suggesting the young age of these objects. The youth of HAEBES has since
been confirmed by multiple studies \citep[e.g.][]{ps91,vda98}.

The analogy with CTTSs motivated Herbig's search for the higher mass HAEBES. In general, HAEBES and
CTTSs have a number of observed features in common, most notably infrared excesses indicating
circumstellar material \citep{bertout88,hillenbrand92,mannings97,natta01,meeus01} and, in the case
of CTTS and some HAEBES, excess UV/optical luminosity attributed to mass accretion onto the star
\citep[e.g.][]{garrison78,basri90,hartigan91,bc95,db11}. Strong emission lines, both permitted and
forbidden, are also common to both groups
\citep{fink85,edwards87,hp92,bc94,hartigan95,cr97,alencar00}. We note that there are currently no
specific criteria differentiating between distinct PMS evolutionary phases for HAEBES as there is
for the TTSs (CTTSs vs weak line TTSs, or WTTSs), although the idea of HAEBES representing an
evolutionary sequence has been outlined in several studies \citep[e.g.][]{cr97,vda97,malfait98}.
Thus we consider HAEBES as a single group even though observed features can vary widely from star to
star and a continuum of evolutionary stages is most likely present within any large HAEBE sample.
Although HAEBES share many observed properties with CTTSs, it is still unclear as to the extent to
which HAEBES interact with their circumstellar environments in ways similar to CTTSs. 
 
The presence of circumstellar disks around many HAEBES is now well established
\citep[e.g.][]{grady00,natta01,dent05,eisner07,matter14}. There is also significant evidence that
many HAEBES are actively accreting material from their disks \citep{db11,mend11,mend13}. The strong
Balmer lines and excess UV/optical continuum emission in CTTSs are generally accepted to be produced
by accreting disk material that falls to the surface along magnetic field lines from a truncation
point in the disk at several stellar radii above the star \citep{bouvier07}. This scenario, termed
magnetospheric accretion (MA), can account for many of the observed properties and spectral features
in CTTSs \citep{edwards94,hartmann94,cg98,muzz98,muzz01}. In addition, strong magnetic fields, which
are required in order for MA to occur, are ubiquitous on TTSs \citep[e.g.][]{jk07}. Magnetospheric
accretion has also been used to explain accretion signatures in some HAEBES \citep{muzz04,grady10}.
\citet{mend11} calculated accretion rates for a large sample of HAEBES by assuming that MA operates
in these systems. 

It is unclear, however, that MA is the dominant accretion mechanism in HAEBES. Perhaps the largest
uncertainty regarding MA in HAEBES is the current lack of detected magnetic fields on these objects.
\citet{wade07} analyzed 50 HAEBES using low--resolution spectropolarimetry and found only 8--12\% of
their sample to have detectable magnetic fields. They also place an upper limit of 300 G on the
longitudinal fields of the undetected sample. More recently, \citet{alecian13} performed a
high--resolution spectropolarimetric study of 70 HAEBES. The high resolution of their data provided
greater sensitivity to Zeeman signatures, enabling detection of weak longitudinal fields. Their
results include the confirmation of 5 magnetic HAEBES while only \textit{one} new object (HD 35929)
is reported as magnetic. Although the magnetic field strengths predicted for MA to operate around
HAEBES are closer to $\sim$a few hundred gauss \citep{wade07}, as opposed to the kilogauss fields
required for low mass stars, the lack of significant detections of \textit{any} magnetic fields on a
large majority of these objects, many of which shows signs of active accretion, suggests a different
accretion mechanism is mediating the transfer of mass from the disk to the star. Since outflows
around young stars are known to be correlated with accretion \citep{hartigan95}, a shift from MA to
a different accretion mechanism may impact the type and strength of outflowing material as well.

The focus of this study is to probe the inner accretion and wind launching regions using the \hei
\AA\hspace{0pt} line diagnostic, which has been successfully employed for similar studies of CTTSs
\citep[e.g.][]{dupree05,EFHK} as well as in mass loss studies of Wolf--Rayet stars \citep{howarth92}
and classical Be stars \citep{groh07}. In this paper we present high spectral resolution
observations of the \hei line for a sample of 56 HAEBES. The \hei diagnostic is an excellent tracer
of outflowing material due to its metastable lower level, i.e. unless the electron density is
sufficiently high for collisional de--excitation to be important, the atoms tend to remain in the
lower 2s$\:^3S$ energy level until a photon is absorbed \citep[see][for details on the atomic
data]{kwan11}. Thus, He atoms excited into the lower energy level have the ability to trace the full
velocity extent of the material. Section 2 provides a brief overview of HAEBE magnetic fields and
how these fields potentially mediate accretion and outflows. In \S 3 we discuss our observations and
data reduction procedures. The line profiles and their morphology classifications are given in \S 3.
The incidence of red and blue--shifted absorption features in our sample is also presented in \S 3.
The line profiles are examined and analyzed in \S 5 within the context of disk or stellar winds and
mass infall. In \S 5 we also elaborate on the implications of our observations for HAEBES as a
whole, as well as discuss scenarios potentially responsible for the differences in morphology
statistics observed between HAEBES and CTTSs. A summary of our findings and conclusions are
presented in \S 6. 

\section{MAGNETIC FIELDS: MAGNETOSPHERIC ACCRETION AND OUTFLOWS}

As mentioned in \S 1, magnetic fields have only been confirmed on a handful of HAEBES
\citep{wade07,alecian13} while kilogauss strength magnetic fields seem to be ubiquitous on CTTSs
\citep{jk07}. \citet{wade07} pointed out that the required surface dipolar field strength for
magnetospheric accretion to occur on a typical HAEBE, according to the MA theories of
\citet{konigl91} and \citet{shu94}, is a few hundred gauss, much lower than the kilogauss
fields required for CTTSs. Although a large majority of HAEBES do not have confirmed magnetic
fields, most of the attempted measurements have been performed using polarimetry. The uncertainties
of these measurements are large, often tens to hundreds of Gauss \citep{wade07,alecian13}. For
polarimetric measurements of magnetic fields on CTTSs the detected longitudinal fields are of the
order $\sim$100 G \citep[e.g.][]{donati10,donati11,donati11a} while Zeeman broadening measurements
reveal strong average fields of $\sim$2 kG \citep{jk07}. For the objects in our sample with measured
$\dot{M}$, $R_*$, $M_*$, and $v$sin$i$ we have calculated the predicted mean surface equatorial
magnetic field strength using the theory of \citet{shu94}. The median field strength for our sample
is 470 G, about a factor of 4 lower than for CTTSs. Thus the longitudinal fields measured via
polarimetry may simply be too low to detect under current observational constraints. Zeeman
broadening measurements for HAEBES would provide better constraints on the surface magnetic field
strengths.  However, the high $v$sin$i$ values of HAEBES and fewer number of photospheric absorption
lines compared to CTTS makes Zeeman broadening detections very difficult. In any case, we cannot
rule out MA on HAEBES based on the lack of detected fields; the field strengths, although large
enough to produce MA flows, may simply be too weak to detect using current methods and
instrumentation.

\citet{muzz04} pointed out that if MA is operating on some HAEBES then the large $v$sin$i$ values
for many stars would force the corotation radii to be very small ($\sim$2 $R_*$), outside of which
accretion cannot take place \citep{shu94}. Small corotation radii, or truncation radii, result in
smaller magnetospheres around HAEBES compared to CTTSs. We note that ``smaller'' in this case refers
to the volume of the MA flow where the flow kinematics are dominated by the magnetic field. Thus
smaller magnetospheres are, in general, required in order for MA to operate around HAEBES. We test
this prediction observationally in \S 5.1.2.

Since smaller corotation radii, and in turn smaller magnetospheres, are expected for HAEBES
experiencing MA, the question arises: can magnetocentrifugal winds be driven by small
magnetospheres? Numerical simulations have shown that CTTS magnetospheres can drive strong outflows
from the magnetosphere--disk interaction region \citep[e.g.][]{kurosawa12,zanni13} and this behavior
is expected from analytic MA theories \citet{shu94,mohanty08}. For a wind being driven along a
magnetic field line, the field enforces corotation out to approximately the Alfv\'{e}n radius,
$R_A$. Thus inside $R_A$ the velocity of the outflowing gas is $v_{g}$($r$)=$v$sin$i$($r/R_*$) if we
take $v$sin$i$ to be the approximate rotation velocity of the field line anchored to star. If we
assume that the escape velocity at $R_A$ is approximately the velocity of the gas at $R_A$, we can
estimate $R_A$/$R_*$=[$2GM_*$/($v$sin$i^2$$R_*$)]$^{1/3}$. We note that here $R_A$/$R_*$ differs
from the corotation radius in the disk, $R_{co}$/$R_*$, by a factor of 2$^{1/3}$. The large
$v$sin$i$ values of HAEBES, and thus the small corotation radii, will generally result in smaller
values of $R_A$ in HAEBES than in CTTSs. We will also investigate the difference in $R_A$ between
HAEBES and CTTSs in \S 4.4.

\section{OBSERVATIONS AND DATA REDUCTION}

Our sample of HAEBES (\autoref{tab:tab1}) consists of 56 objects chosen from the catalogues of
\citet{viera03}; \citet{the94}; and \citet{fink84}. Classification as a HAEBE was the only criteria
for being included in our sample. As a result, our sample covers a wide range of spectral types,
from B0 to F2. Our objects most likely cover a wide range in evolutionary status. The evolutionary
implications of our \hei observations will be discussed in \S 4. While this broad selection criteria
makes it difficult to make specific comparisons between groups with distinct observational
properties, it is sufficient for reaching general conclusions about HAEBES as a whole. Physical
parameters for our sample are given in \autoref{tab:tab2}. With the exception of the radial velocity
for some objects, the \autoref{tab:tab2} values are taken from the literature sources indicated in
column 10.

\capstartfalse
\begin{deluxetable*}{lllcccc}
%\rotate
\tablecaption{Log of \hei observations \label{tab:tab1}}
\tablewidth{0pt}
\tablehead{\colhead{Object ID}&\colhead{Instrument}&\colhead{Telescope$^a$}&\colhead{UT Date}&
\colhead{Integration time$^b$ (s)}&\colhead{Co--added spectra}&\colhead{Final S/N$^c$}\\
\colhead{(1)}&\colhead{(2)}&\colhead{(3)}&\colhead{(4)}&\colhead{(5)}&\colhead{(6)}&\colhead{(7)}}
\tabletypesize{\footnotesize}
\startdata
AB Aur    & GNIRS   & GN & 14--Dec--2012 &  60 & 12 & 160\\
BD+41 3731& Phoenix & KP2.1 & 12--Nov--2013 & 600 & 8 & 15\\ 
BD+61 154 & Phoenix & KP4 & 03--Mar--2013 & 800 & 4 & 15\\ 
BF Ori    & Phoenix & KP4 & 02--Mar--2013 & 800 & 4 & 20\\ 
CQ Tau    & Phoenix & KP4 & 27--Feb--2013 & 600 & 4 & 50\\ 
DW CMa    & Phoenix & KP4 & 02--Mar--2013 & 600 & 8 & 10\\ 
HBC 548   & Phoenix & KP2.1 & 08--Nov--2013 & 600 & 2 & 5\\
HD 114981 & GNIRS   & GN & 16--Jan--2013 & 180 & 8 & 175\\
HD 141569 & Phoenix & KP4 & 27--Feb--2013 & 600 & 6 & 55\\ 
HD 142666 & Phoenix & KP4 & 27--Feb--2013 & 750 & 4 & 30\\ 
HD 144432 & Phoenix & KP4 & 28--Feb--2013 & 600 & 6 & 90\\ 
HD 163296 & Phoenix & KP4 & 28--Feb--2013 & 600 & 2 & 200\\
HD 17081  & Phoenix & KP4 & 03--Mar--2013 & 90  & 2 & 95\\ 
HD 190073 & Phoenix & KP2.1 & 08--Nov--2013 & 600 & 4 & 65\\ 
HD 200775 & Phoenix & KP4 & 27--Feb--2013 & 300 & 2 & 20\\ 
HD 244604 & Phoenix & KP4 & 28--Feb--2013 & 600 & 4 & 20\\ 
HD 250550 & Phoenix & KP4 & 27--Feb--2013 & 700 & 4 & 35\\ 
HD 287823 & Phoenix & KP4 & 02--Mar--2013 & 600 & 4 & 10\\ 
HD 34282  & Phoenix & KP4 & 01--Mar--2013 & 800 & 4 & 10\\ 
HD 34700  & Phoenix & KP4 & 27--Feb--2013 & 600 & 2 & 100\\
HD 35187  & GNIRS   & GN & 14--Dec--2012 & 110 & 2 & 125\\
HD 36408  & GNIRS   & GN & 21--Dec--2012 &  65 & 4 & 290\\
HD 37490  & GNIRS   & GN & 13--Dec--2012 &  30 & 8 & 15\\ 
HD 38120  & Phoenix & KP2.1 & 09--Nov--2013 & 600 & 4 & 25\\ 
HD 50083  & GNIRS   & GN & 13--Dec--2012 &  90 & 8 & 215\\
HD 52721  & GNIRS   & GN & 13--Dec--2012 &  65 & 8 & 120\\
HD 53367  & Phoenix & KP4 & 02--Mar--2013 & 300 & 2 & 80\\ 
HK Ori    & Phoenix & KP2.1 & 08--Nov--2013 & 600 & 4 & 15\\ 
IL Cep    & Phoenix & KP2.1 & 09--Nov--2013 & 600 & 4 & 25\\ 
IP Per    & Phoenix & KP4 & 28--Feb--2013 & 800 & 5 & 15\\ 
IRAS 15462--2551 S & Phoenix & KP4 & 27--Feb--2013 & 800 & 4 & 10\\ 
LkH$\alpha$ 215 & Phoenix & KP4 & 01--Mar--2013 & 800 & 2 & 15\\ 
MWC 1080  & Phoenix & KP2.1 & 10--Nov--2013 & 600 & 8 & 35\\ 
MWC 120   & GNIRS   & GN & 17--Dec--2012 & 140 & 4 & 165\\
MWC 137   & Phoenix & KP4 & 28--Feb--2013 & 600 & 2 & 5\\
MWC 480   & GNIRS   & GN & 13--Dec--2012 & 105 & 8 & 135\\
MWC 610   & Phoenix & KP2.1 & 09--Nov--2013 & 600 & 4 & 50\\ 
MWC 614   & Phoenix & KP2.1 & 10--Nov--2013 & 600 & 4 & 75\\ 
MWC 758   & GNIRS   & GN & 13--Dec--2013 & 145 & 8 & 160\\
MWC 863   & Phoenix & KP4 & 28--Feb--2013 & 600 & 4 & 95\\ 
MWC 953   & Phoenix & KP4 & 28--Feb--2013 & 600 & 2 & 50\\ 
T Ori     & Phoenix & KP4 & 01--Mar--2013 & 800 & 4 & 15\\ 
UX Ori    & Phoenix & KP4 & 27--Feb--2013 & 600 & 4 & 10\\ 
V1185 Tau & Phoenix & KP2.1 & 09--Nov--2013 & 600 & 2 & 10\\ 
V1578 Cyg & Phoenix & KP2.1 & 11--Nov--2013 & 600 & 4 & 15\\ 
V1685 Cyg & Phoenix & KP2.1 & 08--Nov--2013 & 600 & 3 & 20\\ 
V1977 Cyg & Phoenix & KP2.1 & 14--Nov--2013 & 600 & 8 & 10\\ 
V346 Ori  & Phoenix & KP2.1 & 08--Nov--2013 & 600 & 4 & 10\\ 
V351 Ori  & Phoenix & KP4 & 01--Mar--2013 & 800 & 2 & 20\\ 
V374 Cep  & Phoenix & KP2.1 & 09--Nov--2013 & 600 & 6 & 30\\ 
V380 Ori  & Phoenix & KP4 & 03--Mar--2013 & 800 & 2 & 15\\ 
V718 Sco  & Phoenix & KP4 & 27--Feb--2013 & 600 & 4 & 45\\ 
V791 Mon  & Phoenix & KP4 & 03--Mar--2013 & 800 & 4 & 20\\ 
VY Mon    & Phoenix & KP4 & 03--Mar--2013 &1200 & 8 & 5\\  
XY Per    & Phoenix & KP2.1 & 08--Nov--2013 & 600 & 4 & 40\\ 
Z CMa     & GNIRS   & GN & 15--Dec--2012 &  90 & 8 & 130\\
\enddata
\tablenotetext{a}{GN=Gemini North; KP2.1=Kitt Peak 2.1m; KP4=Kitt Peak 4m}
\tablenotetext{b}{Integration time per individual exposure}
\tablenotetext{c}{Continuum S/N for the co--added, un--binned spectrum}
\end{deluxetable*}
\capstarttrue

\capstartfalse
\begin{deluxetable*}{lccccccccc}
\tablecaption{HAEBE sample physical parameters \label{tab:tab2}}
\tablewidth{0pt}
\tablehead{\colhead{} & \colhead{} & \colhead{$v_{rad}$} &
\colhead{$M_*$}&\colhead{$R_*$}&\colhead{\textit{v}sin\textit{i}}&\colhead{log($\dot{M}$)} &
\colhead{Disk} & \colhead{\textit{i}} & \colhead{}\\
\colhead{Object ID}&\colhead{Spectral Type}&\colhead{(km s$^{-1}$)}&
\colhead{($\Msun$)}&\colhead{($\Rsun$)}&\colhead{(km s$^{-1}$)}&
\colhead{(M$_\odot$ yr$^{-1}$)}&\colhead{Detected}&\colhead{($^\circ$)}&\colhead{References$^a$}\\
\colhead{(1)} & \colhead{(2)} & \colhead{(3)} & \colhead{(4)} & \colhead{(5)} &
\colhead{(6)} & \colhead{(7)} & \colhead{(8)} & \colhead{(9)} & \colhead{(10)}}
\tabletypesize{\scriptsize}
\startdata
AB Aur     & A0& 24.7 & 2.50 & 2.62 & 116 &-6.85  & Y &   40  &1,2,7\\
BD+41 3731 & B5& -14.0& 5.50 & 3.80 & 345 &\nodata & N &\nodata&1,8\\
BD+61 154  & B8& -16.0& 3.40 & 2.42 & 112 &\nodata & Y &   70  &1,6,9\\
BF Ori     & A2& 22.0 & 2.58 & 3.26 & 39 &$<$-8.00& Y &\nodata&1,2,6,8\\
CQ Tau     & F2& 35.7 & 2.93 & 5.10 & 98 &$<$-8.30& Y &   29  &1,2,12,37\\
DW CMa$^\dagger$ & B3&\nodata&\nodata&\nodata&\nodata&\nodata& Y &\nodata&4,13\\
HBC 548$^\dagger$& B9&\nodata& 3.80 & 3.20&\nodata&\nodata & Y &\nodata&6\\
HD 114981  & B5& -50.0 & 7.90 & 7.00 & 239 &\nodata & N &\nodata&1\\
HD 141569  & A0& 35.7 & 2.33 & 1.94 & 228 & -6.90  & Y &   55  &1,2,14\\
HD 142666  & A5& -7.0 & 2.15 & 2.82 & 65 & -7.22  & Y &\nodata&1,3,15\\
HD 144432  & A7& -3.0 & 1.95 & 2.59 & 79 &$<$-7.22& Y &$\sim$30&1,2,16,17\\
HD 163296  & A1& -9.0 & 2.23 & 2.28 & 129& -7.16  & Y &\nodata &1,3,6,18\\
HD 17081   & B8& 11.5 & 4.65 & 4.84 & 20 &\nodata & N &\nodata&1,27,33\\
HD 190073  & A1&  0.2 & 2.85 & 3.60 &  4 &\nodata & Y &   45  &1,19\\
HD 200775  & B4& -23.3& 10.70 & 10.4 & 26&\nodata & Y &   55  &1,20\\
HD 244604  & A4& 26.8 & 2.66 & 3.69 & 98& -7.20  & Y &\nodata&1,3,21\\
HD 250550  & B8& -22.0& 4.80 & 3.50 & 79& -7.80  & Y &\nodata&1,3,13\\
HD 287823  & A0& -0.3 & 2.50 & 2.60 & 10&\nodata & ? &\nodata&1\\
HD 34282   & A3& 16.2 & 1.59 & 1.66 & 105&$<$-8.30& Y &\nodata&1,2,18,22\\
HD 34700   & F9& 21.0 & 2.40 & 4.20 & 46& -8.30  & Y &\nodata&2,23\\
HD 35187   & A2& 27.0 & 1.93 & 1.58 & 93& -7.60  & Y &\nodata&1,3,24\\
HD 36408   & B8& 15.0 & 4.10 & 3.50 &\nodata&$<$-8.00& ? &\nodata&2\\
HD 37490   & B4& 21.0 &\nodata&\nodata& 180&\nodata & N &\nodata&6,25,26\\
HD 38120   & B9& 28.0 & 2.49 & 1.91 & 97&-6.90  & Y?&  $<$8 &1,3,35\\
HD 50083   & B4& -0.5 & 12.10 & 10.0 & 233&\nodata & ? &\nodata&1\\
HD 52721   & B3& 21.0 & 9.10 & 5.00 & 215&\nodata & N &\nodata&1,6\\
HD 53367   & B1& 47.2 & 16.10 & 7.10 & 41&$<$-8.92& N &\nodata&1,3,6\\
HK Ori     & A3& 14.4 & 3.00 & 4.10 & 60& -5.24  & Y &\nodata&2,6\\
IL Cep     & B4& -39.0&\nodata&\nodata& 179 &\nodata & ? &\nodata&1\\
IP Per     & A3& 13.7 & 1.86 & 2.10 & 80&\nodata & ? &\nodata&1\\
IRAS 15462-2551 S$^\dagger$& A5&\nodata &\nodata &\nodata &\nodata&\nodata & Y &   90  &4,27\\
LkH$\alpha$ 215 & B7& 0.1 & 5.8 & 5.9 & 210&\nodata & N? &\nodata&1,6,10,13\\
MWC 1080$^\dagger$ & B1&\nodata& 17.4 & 7.3 &\nodata&\nodata & Y &   83  &1,6,10\\
MWC 120    & B9& 47.0 & 3.94 & 4.60 & 120&-6.85  & Y &\nodata&1,3,\\
MWC 137$^\dagger$& B1 &\nodata &\nodata &\nodata &\nodata&\nodata & Y &   80  &6,10,13\\
MWC 480    & A4& 12.9 & 1.93 & 1.93 & 98&$<$-7.23& Y &   37  &1,2,12\\
MWC 610    & B3& 14.0 & 8.00 & 4.70 & 219&\nodata & ? &\nodata&1\\
MWC 614    & A0& 15.1 &\nodata &\nodata &\nodata &-6.59   & Y &\nodata&3,28\\
MWC 758    & A5& 17.8 & 2.90 & 4.40 & 54&-6.05   & Y &   21  &1,3,12\\
MWC 863    & A1& -5.0 & 2.56 & 2.89 & 108&-6.12   & Y &   38  &1,2,29\\
MWC 953    & B3& 23.0 &\nodata &\nodata &\nodata&\nodata & ? &\nodata&5\\
T Ori      & A3& 56.1 & 3.13 & 4.47 &147 &-6.60   & Y &\nodata&1,2,30\\
UX Ori     & A1& 12.0 & 6.72 & 12.1 &221&-7.18   & Y & $<$8  &1,3,6,35\\
V1185 Tau  & A2& 1.0  & 2.04 & 1.75 &250 &\nodata & ? &\nodata&1\\
V1578 Cyg  & A1& -3.0 & 5.90 & 9.70 &199 &\nodata & Y &\nodata&1,6\\
V1685 Cyg$^\dagger$ & B4& -16.0 &\nodata &\nodata &\nodata & \nodata & Y & 41? &30,34\\
V1977 Cyg  & B9& -13.0 & & &\nodata &\nodata & ? &\nodata& 36 \\
V346 Ori   & A7& 20.0 & 1.72 & 1.96 &116&-6.90   & N?&\nodata&1,3,35\\
V351 Ori   & A6& 15.0 & 2.88 & 4.38 & 100&\nodata & ? &\nodata&1\\
V374 Cep$^\dagger$ & B5&\nodata&\nodata& \nodata&\nodata & \nodata & ? &\nodata&\nodata\\
V380 Ori   & B9& 27.5 & 2.87 & 3.00 & 7&-5.60   & Y &\nodata&1,3,6\\
V718 Sco   & A4& -3.6 & 1.93 & 2.25 &113&\nodata & Y?&   32? &1,35\\
V791 Mon   & B5& -2.6 &\nodata &\nodata &\nodata &\nodata & ? &\nodata&4\\
VY Mon$^\dagger$& B8&\nodata &\nodata &\nodata &\nodata&\nodata & Y &   40  &1,9\\
XY Per     & A2& 2.0 & 1.95 & 1.65 & 224&-7.02   & ? &\nodata&3\\
Z CMa      & B9& -27.0 & 3.80 & 3.20 &\nodata&-6.72   & Y &\nodata&3,31\\
\hline
\enddata

\tablenotetext{a}{1=\citet{alecian13}, 2=\citet{mend11}, 3=\citet{db11}, 4=\citet{viera03},
5=\citet{carmona10}, 6=\citet{hillenbrand92}, 7=\citet{tang12}, 8=\citet{fink84}, 9=\citet{boss11},
10=\citet{aa09}, 11=\citet{guillo11}, 12=\citet{guillo13}, 13=\citet{verhoeff12}, 14=\citet{thi14},
15=\citet{schegerer13}, 16=\citet{eisner09}, 17=\citet{chen12}, 18=\citet{marinas11},
19=\citet{ragland12}, 20=\citet{okamoto09}, 21=\citet{vink02}, 22=\citet{natta04},
23=\citet{ackeancker04}, 24=\citet{oud92}, 25=\citet{fuente02}, 26=\citet{mg01},
27=\citet{perrin06}, 28=\citet{liu07}, 29=\citet{fukagawa03}, 30=\citet{eisner04},
31=\citet{schutz05}, 32=\citet{garcia06}, 33=\citet{malfait98}, 34=\citet{hernandez04},
35=\citet{dent05}, 36=\citet{corporon99}, 37=\citet{mend11a}}
\tablecomments{Stars with unknown radial velocities are marked with a $\dagger$.}
\end{deluxetable*}
\capstarttrue

\subsection{Observations}

Our observations were carried out using two instruments on three telescopes: GNIRS (11 objects) on
Gemini North \citep{elias06}, and the Phoenix echelle spectrograph \citep{hinkle98} (45 objects) on
the Mayall 4 m and KPNO 2.1 m. Individual object observations are detailed in \autoref{tab:tab1}.
Although the resolving power of GNIRS (R$\sim$18,000) is much lower than Phoenix (R$\sim$50,000), we
found that almost all of the He I $\lambda$10830 features in our sample are broad and strong enough
to be resolved at the lower resolution. Thus, little to no information is lost in the GNIRS spectra
and similar mass flow scenarios can be investigated using both sets of data. 

The GNIRS data was obtained in queue mode during the 2012B semester using the 0.10''x49'' slit with
the long camera and the 110 lines mm$^{-1}$ grating resulting in a velocity resolution of $\sim$17
km s$^{-1}$. Images were taken in nodded pairs with an offset of $\pm$10''. The X\_G0518 order blocking
filter was employed to isolate a single order around 1.1 $\mu$m providing wavelength coverage from
1.0678--1.0982 $\mu$m. Individual exposure times for the GNIRS sample ranged from 15.0--180.0 s
yielding a typical S/N$\sim$150 for 4--6 co--added exposures. Telluric standards at a similar airmass
were observed immediately before or after each object. Arc lamp exposures were obtained for
wavelength calibration purposes.

The Phoenix observations were obtained during three separate runs in 2013. The 4--pixel slit was
used which corresponds to 0.7''x28'' at the Mayall 4m and 1.4''x56'' at the KPNO 2.1 m. The grating
was configured to provide wavelength coverage from 1.0810--1.0860 $\mu$m and the J9232 order
blocking filter was used to eliminate light from any overlapping orders. This setup yields a
velocity resolution of $\sim$6 km s$^{-1}$ at either telescope. Spectroscopic standards were
obtained for the entire range of HAEBE spectral types; telluric standards were obtained at a variety
of airmasses. ThArNe lamp exposures were taken in order to provide wavelength calibrations.
Signal--to--noise ratios varied significantly for the Phoenix sample depending on the combination of
object brightness and the telescope used for the observation. Almost all 2.1 m targets required
4x900s observations in order to achieve a S/N$>$15, although the faintest objects have S/N$<$10.
Repeating long exposure sequences for individual objects in order to boost the final S/N was avoided
in order to obtain decent S/N exposures of more of the targets in our sample. This strategy was
necessary to obtain quality observations of a large number of HAEBES.  In any case, a low S/N
spectrum is usually adequate for our purposes of identifying a \hei feature and measuring a reliable
equivalent width.  Marginal cases are noted as such in the discussion. 

\begin{figure}\label{fig:fig1}
  \begin{center}
     \includegraphics[scale=.45,trim= 5mm 0mm 0mm 0mm]{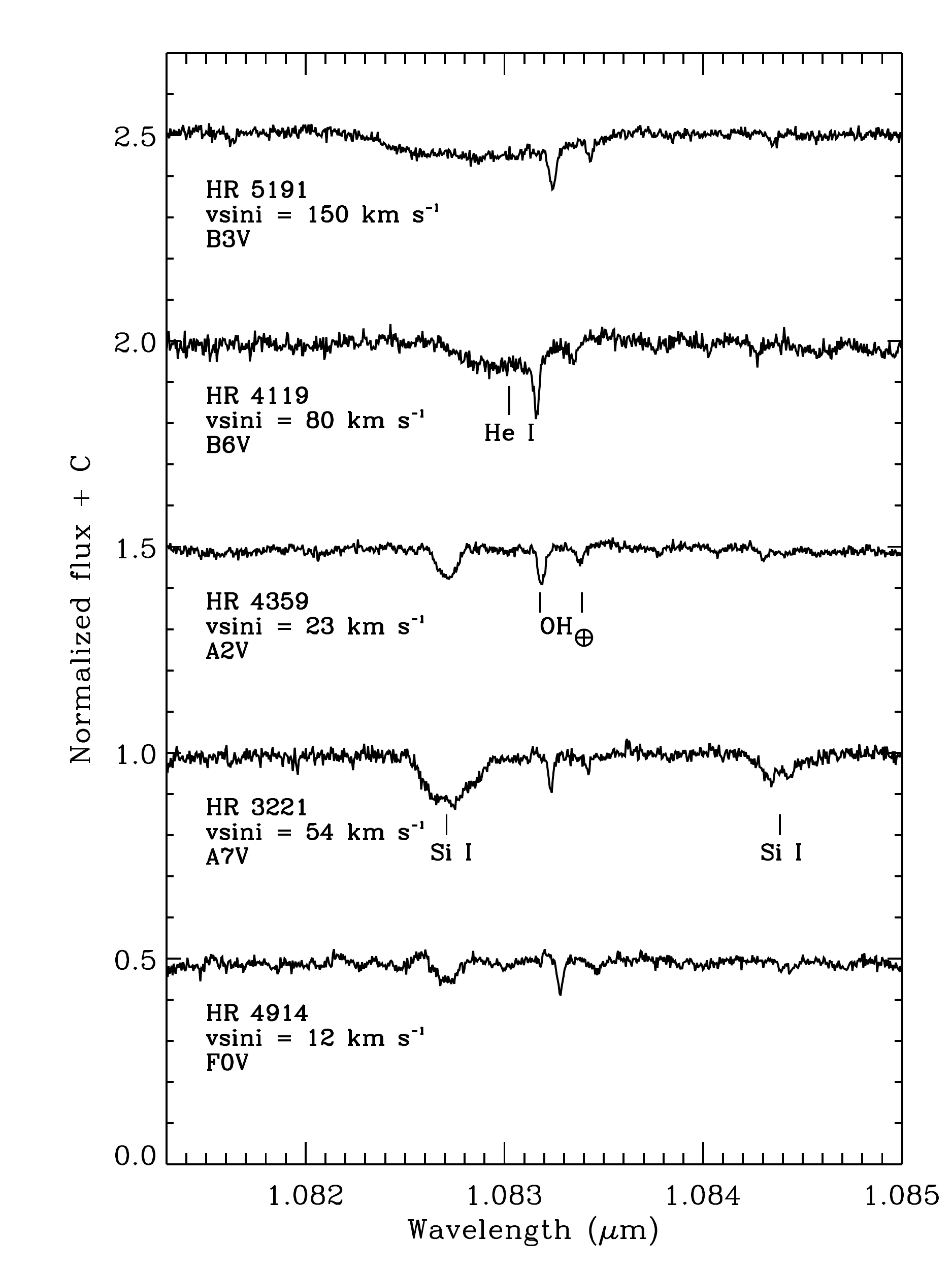}
     \caption{Spectroscopic standard spectra for a sequence of spectral types. OH telluric
absorption lines are marked near the 1.0832 and 1.0834 $\mu$m of the middle spectrum. Stellar
absorption lines are generally weak and are highly doppler--broadened for large values of
$v$sin$i$.}
  \end{center}
\end{figure}

We note that the presence of sub--arcsecond companions, which appear to be common around HAEBES
\citep[e.g.][]{wheelwright10}, has a minimal effect on the analysis presented here. First, the large
flux ratios of the HAEBE primaries to CTTS companions at 10830 \AA\hspace{0pt} ensures that
absorption profiles are dominated by absorption of flux from the primary. Companion CTTS
absorption profiles would be heavily veiled and, with the possible exception of the weak absorption
features seen in a few spectra, would not be in agreement with the absorption strengths observed by
\citet{EFHK} that rarely penetrate below 50\% of the stellar continuum. A similar argument can be
made concerning the strengths of the emission components. Thus any profile contributions from CTTS
companions will likely be small perturbations on top of the primary HAEBE profile. In addition, the
orbital separation of the binary would, in general, have to be very small in order to enable the
absorption of flux from the primary by material emitted by the CTTS companion. Small physical
separations of HAEBE binaries seem to be rare \citep{wheelwright10} so these objects are unlikely to
affect our analysis in this way. 

\subsection{Data Reduction}

All of the data were reduced using custom IDL routines. Each pair of images was differenced and
flat--fielded. The differenced spectra were then optimally extracted and co--added after being
interpolated onto the same wavelength scale. Any remaining bad pixels were manually averaged between
adjacent pixels. For objects observed at high airmass, a telluric standard was scaled and divided
into the object spectrum in order to remove the atmospheric features, most importantly the OH
absorption lines at $\sim$10832 and 10834 \AA. These lines are present in each of the spectra shown
in \autoref{fig:fig1} and are generally weak, even at high airmass. Any residual telluric absorption not
removed by the telluric standard was masked using a quadratic estimate of the stellar spectrum
across the narrow width of the OH telluric line.  Third--order polynomial wavelength solutions were
obtained by fitting the observed ThArNe lamp exposures using line identifications from
\citet{hinkle01}.   

\begin{figure}\label{fig:fig2}
   \begin{center}
      \includegraphics[scale=.65,trim= 100mm 30mm 0mm 0mm]{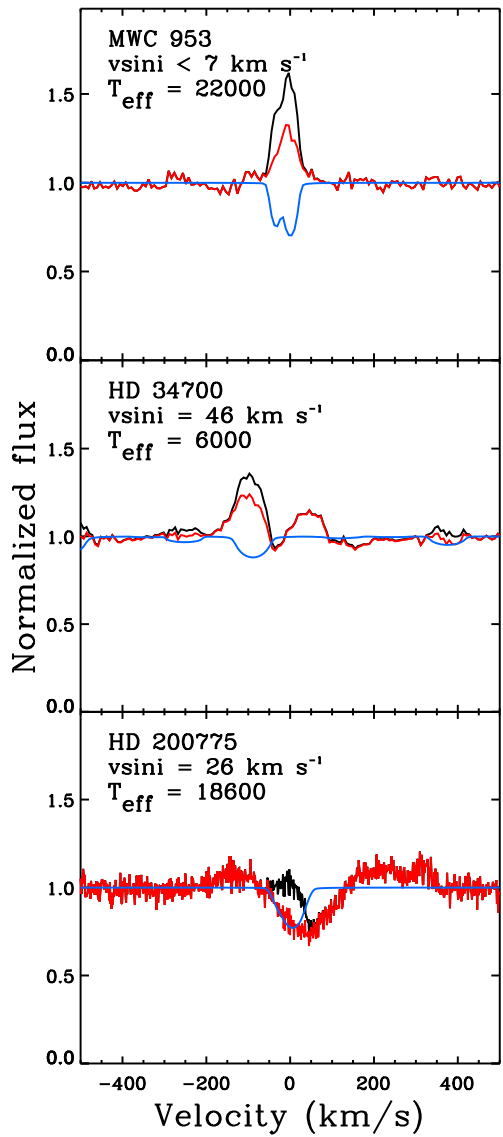}
      \caption{\hei spectra of objects that required the removal of photospheric absorption
 features. The corrected spectrum is plotted in black, the observed spectrum in red, and the
 synthetic photosphere in blue.}
   \end{center}
\end{figure}

In order to evaluate the contamination of photospheric absorption features, we employed both
observed spectroscopic standards and synthetic spectra generated using SYNTHMAG \citep{pisk99}. A
sample of our observed standards are shown in \autoref{fig:fig1}. Many of our observed standards,
which cover spectral types from B2 to F7, do not show any photospheric features. An overwhelming
majority of our science targets also do not show any obvious photospheric lines. This is not
unexpected: according to line lists extracted from VALD \citep{pisk95}, there are few strong
spectral lines (\ltsima 0.90 of the continuum) in our observed wavelength range in A--type stars,
the exception being Si I at $\lambda$10827 and 10843 and a S I line at $\lambda$10821 in the A6--7
objects. Photospheric He I lines only become prominent in $\sim$B5 stars and earlier. Most of our
B--type objects are rotating so fast that the photospheric He I lines are broadened to the point of
becoming negligible contributors to the spectrum. As a check, we visually examined the photospheric
contribution of a broadened synthetic spectrum of similar $T_{eff}$ to each object. Out of the
sample, only HD 34700, MWC 953, and HD 200775 were significantly altered by the subtraction of the
synthetic spectrum. The result of this subtraction is shown in \autoref{fig:fig2}. It can be seen
that the profile morphology is not altered, i.e. the profile classification of these objects does
not change.  The low number of impacted objects is mainly a result of the moderate--to--high
\textit{v}sin\textit{i} values of our targets which tend to make the photospheric contributions weak
compared to the circumstellar features. For the reasons stated above, the large majority of our
sample does not require any removal of photospheric features and we are confident that the spectra
presented in \autoref{fig:fig3} represent true contributions from the circumstellar environment.
First--order normalization was performed using an average of the detector response along the columns
on the detector where the spectra are located. This response function was measured using a median
flat field from each night.  Any residual slope was divided out using a linear fit. 

\section{LINE PROFILES}

\begin{figure*}\label{fig:fig3}
   \begin{center}
      \includegraphics[scale=.90]{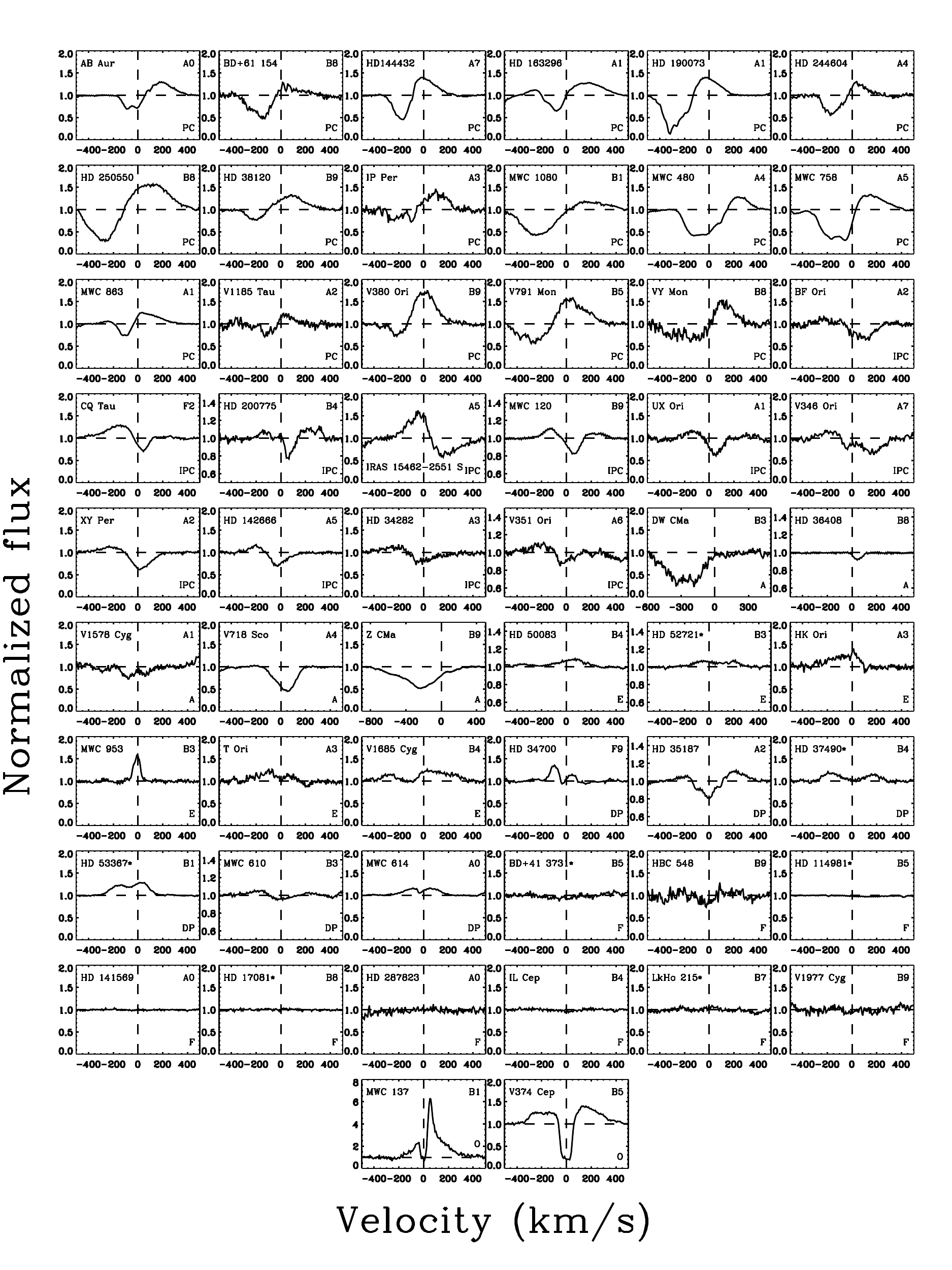}
      \caption{\hei line profiles for the objects listed in \autoref{tab:tab1}. Profiles are grouped according
 to morphology. The morphology abbreviation is given in the lower right hand corner of each plot 
 window. Five--pixel binning is used to plot the objects observed with Phoenix. Spectral types
 are noted in the upper--right of each plot window. Objects that have potentially been misclassified
 as HAEBES (see \S 3.1) are marked with a $\star$ symbol next to their names. Note that the plot 
 range has been narrowed for some objects with weak features.}
   \end{center}
\end{figure*}

\begin{figure*}\label{fig:fig4}
  \begin{center}
     \includegraphics[angle=180,scale=.55,clip,trim= 0mm 0mm 10mm 20mm]{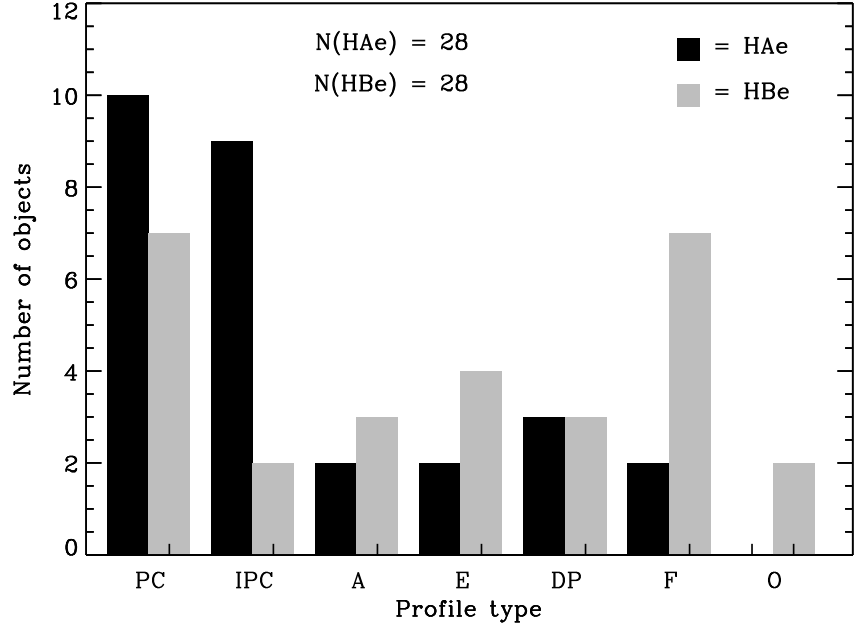}
     \caption{Bar plot of the number of objects in each profile morphology category. The plot labels
are the same as in \autoref{tab:tab3}. This plot includes the potentially misclassified objects from \S 3.1.}
  \end{center}
\end{figure*}

\hei line profiles for our sample are presented in \autoref{fig:fig3} and ordered according to
morphology. To boost S/N and provide better clarity, 5--pixel ($\sim$6.9 km s$^{-1}$) binned spectra
are plotted for the Phoenix observations. A number of measurements were made for each line profile,
and these are provided in \autoref{tab:tab3}. The measurements contained in \autoref{tab:tab3} are
profile classification (column 3), the maximum blue and red--shifted line velocities (cols. 4 and
5), the velocity of peak emission and absorption (cols. 6 and 7), the line fluxes relative to the
continuum at the peak emission and absorption velocities (cols. 8 and 9), and the blue and red
equivalent widths (cols. 10 and 11). In general, our 56 objects can be categorized into six
different morphology groups: 1.  P--Cygni (PC--18 objects); 2. inverse P--Cygni (IPC--10 objects);
3. pure absorption (A--5 objects); 4.  single--peaked emission (E--6 objects); 5. double--peaked
emission (DP--6 objects); and 6.  featureless (F--9 objects). Grouping of the few ambiguous profile
morphologies (e.g. HD 35187, HD 34282) will not have a significant impact on the final morphology
statistics. We note that the profiles of V374 Cep and MWC 137, listed as O for ``other'' in
\autoref{fig:fig3} and \autoref{fig:fig4}, do not neatly fit any of the profile categories given
above. These are the only objects in our sample that display relatively strong, narrow absorption
near the stellar rest velocity. These objects will not be included in the general profile
descriptions of our sample and will be discussed briefly in \S 4.6.  The number of objects in each
morphology group, separated into Ae and Be stars, is shown in \autoref{fig:fig4}. The featureless
spectra will not be discussed in this section. We note that the signal--to--noise of the HBC 547,
V1977 Cyg, and LkH$\alpha$ 215 observations are too low to rule out weak line signatures. For this
same reason we treat them as featureless spectra. This choice does not significantly change the
occurrence statistics of any one group of features. 

\capstartfalse
\begin{deluxetable*}{lcccccccccc}
\tablecaption{HAEBE sample profile parameters \label{tab:tab3}}
\tablewidth{0pt}
\tablehead{\colhead{} & \colhead{} & \colhead{} & \colhead{$V_{max}^{blue}$} &
\colhead{$V_{max}^{red}$} &  \colhead{$V_{peak}^{em}$} & \colhead{$V_{peak}^{abs}$} &
\colhead{$F_{peak}^{em}$} & \colhead{$F_{peak}^{abs}$} & \colhead{W$_{blue}$} &
\colhead{W$_{red}$}\\
\colhead{Object ID}&\colhead{Spectral Type}&\colhead{Profile Type$^a$}&
\colhead{(km s$^{-1}$)}&\colhead{(km s$^{-1}$)}&\colhead{(km s$^{-1}$)}&\colhead{(km s$^{-1}$)}&
\colhead{($F_{cont}$)}&\colhead{($F_{cont}$)}&\colhead{(\AA)}&\colhead{(\AA)}\\
\colhead{(1)} & \colhead{(2)} & \colhead{(3)} & \colhead{(4)} & \colhead{(5)} &
\colhead{(6)} & \colhead{(7)} & \colhead{(8)} & \colhead{(9)} & \colhead{(10)} &
\colhead{(11)}}
\tabletypesize{\scriptsize}
\startdata
AB Aur     & A0& PC  &  -175   &   405   &   185   &   -90   &   1.29  &    0.69 & 1.30 & -1.70 \\
BD+41 3731 & B5&F   & \nodata & \nodata & \nodata & \nodata & \nodata & \nodata & 0.34 &  .27  \\
BD+61 154  & B8&PC  &  -350   &   300   &    25   &   -145  &   1.32  &    0.50 & 3.15 & -1.19 \\
BF Ori     & A2&IPC &  -385   &   285   &  -275   &   110   &   1.07  &    0.68 & 0.23 &  2.03 \\
CQ Tau     & F2&IPC &  -490   &   130   &  -145   &   40    &   1.28  &    0.72 & -2.24&  0.78 \\
DW CMa$^\dagger$ & B3&A   &  -590   &   2     & \nodata &   -255  & \nodata &    0.35 & 8.92 &$\sim$0\\
HBC 548$^\dagger$& B9&F   & \nodata & \nodata & \nodata & \nodata & \nodata & \nodata & 0.70 & -0.20\\
HD 114981  & B5&F   & \nodata & \nodata & \nodata & \nodata & \nodata & \nodata & 0.07 &  0.28 \\
HD 141569  & A0&F   & \nodata & \nodata & \nodata & \nodata & \nodata & \nodata & 0.04 &  0.07 \\
HD 142666  & A5&IPC &  -310   &   150   &  -205   &   -35   &   1.15  &    0.70 & 0.28 &  0.66 \\
HD 144432  & A7&PC  &  -325   &   240   &  -10    &   -175  &   1.40  &    0.46 & 1.59 & -1.58 \\
HD 163296  & A1&PC  &  -355   &   450   &   180   &   -80   &   1.27  &    0.61 & 1.35 & -2.39 \\
HD 17081   & B8&F   & \nodata & \nodata & \nodata & \nodata & \nodata & \nodata & 0.17 &  0.04 \\
HD 190073  & A1&PC  &  -480   &   180   &   -30   &  -320   &   1.40  &    0.12 & 4.00 & -1.36 \\
HD 200775  & B4&IPC &  -200   &   355   &   -135  &   40    &   1.07  &    0.73 & 0.01 &  0.29 \\
HD 244604  & A4&PC  &  -290   &   260   &    35   &  -175   &   1.29  &    0.57 & 2.19 & -1.25 \\
HD 250550  & B8&PC  &  -490   &   455   &    120  &  -275   &   1.56  &    0.30 & 4.70 & -5.57 \\
HD 287823  & A0&F   & \nodata & \nodata & \nodata & \nodata & \nodata & \nodata & 0.12 &  0.24 \\
HD 34282   & A3&IPC &  -300   &   300   &   -250  &$\sim$0  &   1.09  &    0.75 & 0.58 &  1.54 \\
HD 34700   & F9&DP  &  -175   &   100   &    5    & \nodata &   1.22  & \nodata & -0.61& -0.14 \\
HD 35187   & A2&DP  &  -350   &   390   &   205   &   10    &   1.11  &    0.81 & 0.26 & -0.17 \\
HD 36408   & B8&A   &  -20    &   110   & \nodata &   45    & \nodata &    0.92 & 0.03 &  0.24\\
HD 37490   & B4&DP  &  -305   &   290   &   -170  & \nodata &   1.17  & \nodata & -1.04& -0.71 \\
HD 38120   & B9&PC  &  -325   &   365   &   90    &  -210   &   1.31  &    0.78 & 0.77 & -2.31 \\
HD 50083   & B4&E   &  -190   &   290   &   70    & \nodata &   1.08  & \nodata & -0.25& -0.48 \\
HD 52721   & B3&E   &  -185   &   310   &   -65   & \nodata &   1.06  & \nodata & -0.34& -0.37 \\
HD 53367   & B1&DP  &  -300   &   165   &   30    & \nodata &   1.28  & \nodata & -1.86& -1.04 \\
HK Ori     & A3&E   &  -310   &   120   &   30    & \nodata &   1.40  & \nodata & -1.71& -0.75 \\
IL Cep     & B4&F   & \nodata & \nodata & \nodata & \nodata & \nodata & \nodata & 0.06 &  0.06 \\
IP Per     & A3&PC  &  -310   &   260   &   100   &  -240   &   1.44  &    0.77 & 1.24 & -2.37 \\
IRAS 15462-2551 S$^\dagger$& A5&IPC & -225 & 430 &  -45    &   162   &   1.58  &    0.57 & -2.83& 2.50 \\
LkH$\alpha$ 215 & B7&F & \nodata & \nodata & \nodata & \nodata & \nodata & \nodata & -0.07 & -0.07\\
MWC 1080$^\dagger$ & B1&PC  &  -530   &   465   &   150   &  -245   &   1.18  &    0.43 &  6.10& -1.68 \\
MWC 120    & B9&IPC &  -245   &   370   &  -135   &   65    &   1.10  &    0.83 & -0.42&  0.32 \\
MWC 137$^\dagger$& \nodata&\nodata&-180&550   &   100   &   55    &   6.48  &    0.57 & -2.24&-18.8\\
MWC 480    & A4&PC  &  -275   &   405   &   235   &  -125   &   1.27  &    0.42 &  4.25& -0.07 \\
MWC 610    & B3&DP  &  -275   &   160   &   -185  &  -10    &   1.03  &    0.94 & -0.05&  0.16 \\
MWC 614    & A0&DP  &  -250   &   230   &   -90   & \nodata &   1.16  & \nodata & -0.87& -0.79\\
MWC 758    & A5&PC  &  -360   &   415   &   145   &   -60   &   1.33  &    0.31 &  6.00& -2.38 \\
MWC 863    & A1&PC  &  -330   &   330   &   35    &  -105   &   1.25  &    0.73 &  0.63& -1.65 \\
MWC 953    & B3&E   &  -65    &   45    &   -5    & \nodata &   1.74  & \nodata & -0.98& -0.48\\
T Ori      & A3&E   &  -260   &   195   &   -80   & \nodata &   1.23  & \nodata & -1.17& -0.46\\
UX Ori     & A1&IPC & -230    &   150   &   -135  &   55    &   1.19  &    0.65 & -0.54&  1.34\\
V1185 Tau  & A2&PC  & -170    &   190   &   40    &   -120  &   1.23  &    0.74 &  0.92& -0.78\\
V1578 Cyg  & A1&A   & -170    &   120   &   -80   & \nodata & \nodata &    0.82 &  0.98&  0.64\\
V1685 Cyg$^\dagger$ & B4&E   & -80     &   355   &   35    & \nodata &   1.25  & \nodata & -0.30&-2.01\\
V1977 Cyg  & B9&F   & \nodata & \nodata & \nodata & \nodata & \nodata & \nodata &  0.14&  0.25\\
V346 Ori   & A7&IPC & -300    &   320   &  -175   &  170    &   1.13  &    0.67 & -0.28&  0.67\\
V351 Ori   & A6&PC  & -325    &   280   &  -185   &  -15    &   1.07  &    0.87 & -0.22&  0.17 \\
V374 Cep$^\dagger$ & B5&\nodata & -350 &   445   &   145   &   10    &   1.40  &    0.19 & -0.77& -1.68 \\
V380 Ori   & B9&PC  & -220    &   245   &   5     &  -220   &   1.68  &    0.75 & -0.94& -3.05 \\
V718 Sco   & A4&A   & -155    &   205   & \nodata &   50    & \nodata &    0.46 &  1.07&  2.49\\
V791 Mon   & B5&PC  & -480    &   320   &   20    &  -280   &   1.54  &    0.60 &  1.98& -3.42\\
VY Mon$^\dagger$& B8&PC  & -445&  340   &   90    &  -140   &   1.51  &    0.65 &  3.63& -2.87\\
XY Per     & A2&IPC & -390    &   205   &  -230   &   20    &   1.12  &    0.62 & -0.10&  1.60\\
Z CMa      & B9&A   & -790    &   250   & \nodata &  -240   & \nodata &    0.51 &  7.46&  0.89\\
\hline
\enddata
\tablenotetext{a}{PC=P--Cygni; IPC=inverse P--Cygni; A=pure absorption; E=pure emission;
DP=double--peaked; F=featureless; O=other}
\tablecomments{Stars with unknown radial velocities are marked with a $\dagger$.}
\end{deluxetable*}
\capstarttrue

The red and blue--shifted absorption statistics for our sample, as well as those for the CTTS study
of \citet{EFHK}\defcitealias{EFHK}{EFHK}, hereafter EFHK, are given in \autoref{tab:tab4}. The 68\%
confidence intervals are calculated using Wilson's score interval \citep{wilson27}. The given
percentage is the observed incidence and not the adjusted estimate given by Wilson's test. We also
employ contingency tests \citep[see][]{feigelson12} to compare the number of red and blue absorption
profiles observed in each group. Due to the small number statistics in most cases, we employ
Fisher's exact test for calculating the \textit{p}--values. The null hypothesis in this case is that
red--shifted and blue--shifted absorption features are equally as likely in both groups of objects.
The \textit{p}--value gives the probability of obtaining the observed distribution of profiles
morphologies given the null hypothesis is true. Thus lower $p$--values indicate higher confidence in
rejecting the null hypothesis. \autoref{tab:tab5} displays the contingency tables and the results of
the Fisher tests are shown in \autoref{tab:tab6}.

\capstartfalse
\begin{deluxetable}{lccccc}
\tablecaption{Statistical summary of absorption features \label{tab:tab4}}
\tablewidth{0pt}
\tablehead{\colhead{}&\multicolumn{3}{c}{This study}&\colhead{}&\colhead{EFHK}\\
           \cline{2-4} \cline{6-6}\\
           \colhead{}&\colhead{HAe$^a$}&\colhead{HBe}&\colhead{Total}&\colhead{}&\colhead{CTTSs}\\
           \colhead{}&\colhead{(1)}&\colhead{(2)}&\colhead{(3)}&\colhead{}&\colhead{(4)}}
\tabletypesize{\footnotesize}
\startdata
Red absorption$^b$ & 36\%$^{+9}_{-8}$ & 15\%$^{+10}_{-6}$ & 27\%$^{+7}_{-6}$ && 46\%$^{+8}_{-8}$\\
Blue absorption    & 36\%$^{+9}_{-8}$ & 45\%$^{+11}_{-10}$ & 40\%$^{+7}_{-7}$ && 69\%$^{+7}_{-8}$\\
Niether$^c$        & 28\%$^{+9}_{-8}$ & 40\%$^{+10}_{-10}$ & 33\%$^{+7}_{-6}$ && 10\%$^{+6}_{-4}$\\
\hline
\enddata
\tablenotetext{a}{Percentage of HAe and HBe stars of the total number in each group excluding the
objects
 from \S 3.1. The percentage in column (3) is the percentage of the total number of
HAEBES (HAe and HBe objects) in the sample.}
\tablenotetext{b}{Note that the absorption percentages include the pure absorption profiles in
addition to the P--Cygni and Inverse P--Cygni profiles.}
\tablenotetext{c}{This includes centered absorption with no apparent velocity shift, pure emission
line profiles, double--peaked, and featureless spectra.}
\end{deluxetable}
\capstarttrue

The equivalent widths (EW) in columns (10) and (11) in \autoref{tab:tab3} are calculated using the unbinned
spectra. An estimate of the uncertainties in the equivalent widths can be gained by looking at the
EW measurements for the featureless spectra. Typical EW uncertainties range from $\sim$2.0
\AA\hspace{0pt} for spectra with S/N$\sim$10 down to $\sim$0.1 \AA\hspace{0pt} for spectra with
S/N$\sim$200. Radial velocity measurements (column 3 in \autoref{tab:tab2}) for most objects were taken from
the literature. For objects which do not have previously determined radial velocities, we used
shifted and rotationally broadened synthetic spectrum fits to high resolution optical spectra (data
collected separately, Cauley \& Johns--Krull 2014, in prep) to provide an estimate. For objects
without identifiable photospheric features, we attempted gaussian fits to the wings of the H$\alpha$
emission profile. This method proved to be unreliable and these objects are analyzed without
accounting for radial velocity. The 8 objects for which we do not have reliable RV estimates are
indicated in \autoref{tab:tab2} and \autoref{tab:tab3}. Interpretations of these object profiles are not significantly
affected, assuming the RV values are not unreasonably high ($>\lvert 100\rvert$ km s$^{-1}$), which
we do not see in the stars with known RVs.

Due to the importance of generating accurate statistics concerning the occurrence of mass accretion
and wind flows around HAEBES, we discuss below (\S 4.1) the potential inclusion of objects in our
sample that are not appropriate for comparison to CTTSs.  It is important to note that although
objects may be young (i.e.  pre--main sequence) and thus correctly classified as a HAEBE, their
inactivity or lack of interaction with their environments renders them irrelevant for comparing the
methods by which HAEBES and CTTSs evolve. This is due to the fact that all CTTSs are accreting from
their disks and the interaction of the star and disk is what produces the characteristic CTTS
signatures. Thus HAEBES that are not surrounded by a close circumstellar accretion disk, and thus
are non--accreting, are more similar to WTTSs, although there is currently no precise distinction
between HAEBES with active disks and those without. Objects of this nature in our sample are
excluded from most of the analysis. In sections 4.2--4.5 brief interpretations of the line profiles
will be given along with the morphology descriptions. We will elaborate on the nature of the blue
and red--shifted absorption profiles in \S 5.1.          

\subsection{Potentially misclassified objects}

Based on the line profiles shown here and separately obtained H$\alpha$ profiles, as well as
circumstellar disk indicators from the literature, we have identified 7 objects that may not be
pre--main sequence stars.  We note that classical Be stars often show featureless spectra or
emission at \hei and the emission is often double--peaked \citep{groh07}. Discussions of the line
profile morphology statistics will take these potential interlopers into account throughout this
section and \S 5. There are two objects in our sample, HBC 548 and HD 141569, that have featureless
spectra but also show evidence for circumstellar disks. HD 141569 appears to have significant grain
growth in its disk and there is tentative evidence that a massive planet is present \citep{thi14}.
It also shows significant evidence of accretion \citep{mend11}. Thus although this object shows no
evidence of activity at $\lambda$10830, its young age ($\sim$5 Myr) and the direct detection of warm
molecular gas in its disk does not allow us to rule out its classification as a HAEBE. There is
little information in the literature concerning the nature of the circumstellar material surrounding
HBC 548 so we maintain its status as a HAEBE for our analysis. Objects not discussed in this section
are assumed to be HAEBES that are interacting with their circumstellar environments. 

\subsubsection{BD+41 3731}

This object has a featureless \hei spectrum. In addition, it shows a purely photospheric H$\alpha$
profile and no sign of a disk has been detected. \citet{the94} in fact rejected this object as a
HAEBE and \citet{fink84} also find little evidence for classification as a pre--main sequence
object. Our observations add evidence to support its non--PMS nature and we suggest that it not be
included as such in future studies.

\subsubsection{HD 114981}

HD 114981 was labeled as a HAEBE candidate by \citet{viera03} due to its H$\alpha$ emission and its
detection in the IRAS Faint Source Catalogue. The spectrum at \hei is featureless and the H$\alpha$
profile shows a very symmetric, broad, double--peaked emission profile that is more similar to those
found in rotating classical Be star disks than in the circumstellar environments of HAEBEs. It also
has a very small $H-K$ excess ($\sim$0.03) indicating a lack of inner disk material. We suggest that
this object is actually a classical Be star. 

\subsubsection{HD 17081}

This star shows no features at He I $\lambda$10830. \citet{dent05} did not detect CO emission from HD
17081, indicating that it lacks a gaseous disk. In addition, the observations of \citet{malfait98}
show that its IR excess begins at $\sim$12 $\mu$m making it closer to a Vega--type object that has
completely dissipated the gas component of the disk and is left primarily with debris. Thus HD 17081
has most likely evolved past the point of accreting material from its surroundings, and probably
lacks any significant close circumstellar disk.   

\subsubsection{HD 37490}

This object shows no signs of a disk \citep{mg01} nor molecular material in its near vicinity
\citep{fuente02}. The very symmetric double--peak at \hei matches the double--peaked H$\alpha$
profile. These line morphologies are very similar to those of a classical Be star, which is more
likely the true nature of this object.

\subsubsection{HD 52721}

\citet{hillenbrand92} find no evidence of an IR excess around this object. Our observations show
very weak, broad emission at He I $\lambda$10830; the H$\alpha$ emission profile is single--peaked,
broad, and centered at the stellar rest velocity. The low \textit{v}sin\textit{i} (21 km s$^{-1}$)
suggests a small viewing angle which is consistent with single--peaked, centered emission at
H$\alpha$ if the emission arises in classical Be disk. Thus this object is more likely a classical
Be star than a HAEBE.

\subsubsection{HD 53367}

No IR excess was detected by \citet{hillenbrand92} around HD 53367. Furthermore, \citet{pogodin06}
find significant evidence of a $\sim$20 $\Msun$ primary object in orbit with a $\sim$4--5 $\Msun$
secondary star. They conclude that the massive primary has already evolved onto (or beyond) the main
sequence. By chance, we collected two spectra of this object separated by 9 months. The
double--peaked nature of the emission line is present in both spectra with clear evidence that the
peaks have shifted in velocity. Finally, the H$\alpha$ profile is well matched by two broad,
relatively weak gaussian profiles suggesting two separate emission lines centered at different
velocities. Thus our observations support the close binary scenario, indicating that the emission
lines may not be an indicator of the age of the primary.     

\subsubsection{LkH$\alpha$ 215}

There is mixed evidence for a disk surrounding LkH$\alpha$ 215. \citet{hillenbrand92} find a
substantial IR excess past 1 $\mu$m that is typical of their Class I objects. \citet{verhoeff12},
however, using higher resolution IR imaging, find that most of the IR flux is associated with the
surrounding nebula and suggest that this object may instead be a classical Be star. No disk is
detected at millimeter wavelengths by \citet{aa09}. Our featureless \hei spectrum is more consistent
with the classical Be star scenario.

Given the uncertain nature of the 7 objects discussed above, we do not include them in the
discussion below. We also recommend that they be excluded from any future HAEBE studies and that
their classification be re--examined. 

\capstartfalse
\begin{deluxetable*}{ccccccccc}
\tablecaption{Contingency data \label{tab:tab5}}
\tablewidth{0pt}
\tablehead{\colhead{}&\colhead{}&\colhead{\bf{HAe}}&\colhead{\bf{HBe}}&\colhead{\bf{Total}}&
           \colhead{}&\colhead{\bf{HAEBES}}&\colhead{\bf{CTTS}}&\colhead{\bf{Total}}}
\tabletypesize{\footnotesize}
\startdata
\bf{Red}         & Yes   & 10 & 3  & 13 && 13 & 18 & 30 \rule{0pt}{2.2ex}\\
\bf{absorption?} & No    & 18 & 17 & 35 && 35 & 21 & 57 \rule[-1.2ex]{0pt}{0pt}\\ \hline
%                 & Total & 28 & 20 & 48 &&& 48 & 39 & 87 \\
\bf{Blue}        & Yes   & 10 & 9  & 19 && 19 & 27 & 46 \rule{0pt}{2.2ex}\\
\bf{absorption?} & No    & 18 & 11 & 29 && 29 & 12 & 41 \rule[-1.2ex]{0pt}{0pt}\\ \hline
                 & \bf{Total} & 28 & 20 & 48 && 48 & 39 & 87 \rule{0pt}{2.5ex}\\
                 &            &         &&     &&    &    &                 \\
                 &       & \textbf{HAe} & \textbf{CTTS} & \textbf{Total} && \textbf{HBe} &
\textbf{CTTS} & \textbf{Total} \\
\bf{Red}         & Yes   & 10 & 18 & 28 && 3  & 18 & 21 \rule{0pt}{2.2ex}\\
\bf{absorption?} & No    & 18 & 21 & 39 && 17 & 21 & 38 \rule[-1.2ex]{0pt}{0pt}\\ \hline
%                 & Total & 28 & 20 & 48 &&& 48 & 39 & 87 \\
\bf{Blue}        & Yes   & 10 & 27 & 37 && 9  & 27 & 36 \rule{0pt}{2.2ex}\\
\bf{absorption?} & No    & 18 & 12 & 30 && 11 & 12 & 23 \rule[-1.2ex]{0pt}{0pt}\\ \hline
                 & \bf{Total} & 28 & 39 & 67 && 20 & 39 & 59 \rule{0pt}{2.5ex}\\
\enddata
\end{deluxetable*}
\capstarttrue

\capstartfalse
\begin{deluxetable}{llc}
\tablecaption{Contingency test $p$--values \label{tab:tab6}}
\tablewidth{0pt}
\tablehead{\colhead{Feature}&\colhead{Groups}&\colhead{$p$--value}}
\tabletypesize{\footnotesize}
\startdata
Red absorption & HAe vs. HBe     & 0.188 \\%& 3.08 & 79.5\% \\
               & HAEBES vs. CTTSs& 0.076 \\%& 0.44 & 89.5\% \\
               & HAe vs. CTTSs   & 0.457 \\
               & HBe vs. CTTSs   & 0.023 \\
Blue absorption& HAe vs. HBe     & 0.561 \\%& 0.68 & 27.5\% \\
               & HAEBES vs. CTTSs& 0.009 \\%& 0.30 & 99.0\% \\
               & HAe vs. CTTSs   & 0.012 \\
               & HBe vs. CTTSs   & 0.094 \\
\enddata
%\tablenotetext{a}{Confidence that the true odds ratio is not equal to 1.0}
\end{deluxetable}
\capstarttrue

\subsection{P--Cygni profiles and blue absorption}

Objects with P--Cygni (PC) profiles or blue absorption comprise 40\%$^{+7}_{-7}$ of our corrected
sample. The Herbig Ae stars (HAe) show a 36\%$^{+9}_{-8}$ blue--shifted absorption incidence and the
Herbig Be stars (HBe) show a 45\%$^{+11}_{-10}$ incidence. \autoref{tab:tab6} shows there is a significant
difference between the blue absorption incidence between CTTSs and both HAe and HBe stars. The
blue--shifted absorption incidence in HAEBES as a whole is also significantly different from that in
CTTSs. There is no signficant difference in the blue absorption incidence between HAe and HBe stars. 

There are two objects (DW CMa and Z CMa) with strong blue absorption and no emission (classified as
A type profiles). PC line profiles and broad blue--shifted absorption are classical indicators of
outflows and are commonly found in the Balmer lines of many HAEBES \citep[e.g.][]{fink84,viera03}.
The blue absorption in many of our profiles extends out to terminal velocities near -400 -- -500 km
s$^{-1}$ with maximum depth velocities near $\sim$-200 km s$^{-1}$. The maximum blue velocities for
DW CMa and Z CMa are much higher at $\sim$-600 and -800 km s$^{-1}$, respectively. Z CMa is a known
binary consisting of a HAEBE and an FU Orionis--like object \citep[e.g.][]{hinkley13}; DW CMa is an
early B--star \citep{verhoeff12}. Both of these systems have strong wind signatures in the optical
\citep{fink84}.  The strong, extended \hei lines shown here confirm the strong outflows emerging
from both systems.  Four objects (AB Aur, HD 36112, IP Per, and MWC 758) show evidence of an
emission bump near the maximum depth of the absorption profile. 

There are two general types of PC profiles exhibited by our sample: I -- (hereafter PCI) the peak of
the emission component is red--shifted by $>$50 km s$^{-1}$ (e.g. AB Aur, MWC 480), and II -- 
(hereafter PCII) the peak of the emission component is near (within 50 km s$^{-1}$) zero velocity
(e.g. V380 Ori, HD 144432). Profiles PCI and PCII are split almost evenly across the PC objects. The
equivalent widths of the absorption and emission components are approximately equal in 11 of 17
objects, while the absorption equivalent width is substantially larger in 5 of 17 profiles. The
exception is V380 Ori for which the zero velocity--centered emission is much stronger than the
blue--shifted absorption.

We note that the blue absorption signatures exhibited by the objects in this section are very broad
and are more consistent with a stellar wind--like outflows than a broadened inner disk wind. Inner
disk winds tend to produce narrower blue--shifted absorption profiles with relatively weak emission
compared with stellar wind profiles \citep{KEF}. In fact, our HAEBE sample shows no strong evidence
of narrow, blue--shifted absorption indicating a lack of inner--disk winds in our sample. The larger
rotational velocities expected for inner disk winds in HAEBES most likely cannot account for the
observed broad, stellar wind--like profiles due to the narrow range of velocities projected against
the stellar disk. Furthermore, disk winds have trouble reproducing the standard P-Cygni profiles
observed in most of our PC objects \citep{KEF}. One example supporting stellar winds over broadened
disk winds in our sample is HD 190073 which has a $v$sin$i$ of $\sim$4 km s$^{-1}$ and an
inclination angle of $\sim$45$^\circ$, ruling out rotational broadening as the mechanism responsible
for the broad absorption profile. Interpretations of the blue--shifted absorption profiles are given
in \S 5.1.1.  We will return to the lack of disk winds in our sample in \S 5.2.1.

\subsection{Inverse P--Cygni profiles and red absorption}   

Inverse P--Cygni (IPC) profiles are formed through the scattering or absorption of photons by
infalling material, producing red--shifted absorption features which are most easily detected when
the absorption extends below the continuum. Our sample contains 13 objects, or 27\%$^{+7}_{-6}$
(\autoref{tab:tab4}) of the sample, that display IPC characteristics or red--shifted absorption (classified as
"A" type profiles) in their \hei profiles (e.g. CQ Tau, MWC 120). Interpretations of the
red--shifted absorption profiles described here are presented in \S 5.1.2. Of these 13 objects, 10
are HAe stars and 3 are HBe stars. This corresponds to an incidence of 36\%$^{+9}_{-8}$ for the HAe
objects and 15\%$^{+10}_{-6}$ for the HBe objects, indicating that MA is more common among HAe stars
than HBe stars. A contingency test, however, shows (\autoref{tab:tab6}) that these percentages are not
significantly different given the sample sizes. There is also no significant difference in the red
absorption incidence between HAe stars and CTTSs. HBe stars show a significantly lower incidence
than CTTSs. As a whole, the incidence of red--shifted absorption in HAEBES is significantly
different from that in CTTSs. 

The red--shifted absorption typically extends to $\sim$+200-250 km s$^{-1}$. The
notable exception is IRAS 15462-2551 which displays broad, tapering absorption out to $\sim$+500 km
s$^{-1}$. HD 34282 may have absorption that also extends to higher velocities but the continuum is
noisy and difficult to locate. The red--shifted absorption in the profile of MWC 120 is superimposed
on broad, weak emission that extends from -200 km s$^{-1}$ to +300 km s$^{-1}$. Similar behavior is
exhibited by HD 200775, although the noise makes the extent of the emission less clear. Both of
these objects have B spectral types. 

Two of the 13 objects show only red--shifted absorption with no emission: HD 36408 and V718 Sco. HD
36408 shows weak, red--shifted absorption centered at +45 km s$^{-1}$. This signature may be due to
a weak MA flow with a low surface filling factor viewed at a small inclination angle. This would be
consistent with its relatively low accretion rate \citep{mend11}. The profile of V718 Sco shows
deep, asymmetric, slightly red--shifted absorption with relatively steep wings extending to -150 and
+200 km s$^{-1}$. Evidence for accretion onto V718 Sco has been found in its Balmer lines
\citep{guim06} but no estimate exists for the accretion rate.

HD 142666, HD 34282, and V351 Ori all show IPC--type profiles but the maximum absorption is at
slightly blue--shifted wavelengths. We include these objects in the IPC category since their
emission components are clearly blue--shifted and the absorption is only marginally blue--shifted.
Radial velocity estimates for these objects are accurate and cannot explain the blue--shifted
maximum absorption depths. The profiles are entirely circumstellar since all of these objects are
mid A--type stars and have no photospheric features near 10830 \AA. These objects should be
considered marginal IPC morphologies and follow up observations should be conducted to verify the
line shapes.

\subsection{Single--peaked emission}        

The six single--peaked emission objects (10\%) in our sample display a variety of profile
morphologies. HD 50083 and HD 52721 each show very weak, broad emission. These profiles may be
evidence of a weak disk--wind viewed nearly edge on. This scenario is consistent with previous
emission line studies of both objects and their high \textit{v}sin\textit{i} values suggest a high
viewing angle. In addition, the S/N in these observations is high making the emission signature
unambiguous. MWC 953 displays strong, narrow emission centered near the stellar rest velocity. We
note that this object was divided by a B3 (T$_{eff}$$\sim$18,000 K) synthetic spectrum due to the
strong He absorption at 10830 \AA; the original profile was similar in shape but less prominent (see
\autoref{fig:fig2}). The profile of MWC 953 is most readily reproduced by the nearly edge--on polar
stellar wind models of \citet{KEF}. However, there is currently no evidence of an accretion disk
around this object which is believed necessary for the formation of a polar wind. HK Ori shows
broad, tapering emission in the blue out to $\sim$-310 km s$^{-1}$, with a much steeper red wing
terminating at $\sim$115 km s$^{-1}$. Although the peak appears to be slightly red--shifted, the
spectrum is rather noisy and thus the true peak velocity is uncertain. The emission of V1685 Cyg
tapers to $\sim$350 km s$^{-1}$ and may have a second emission peak near -300 km s$^{-1}$, although
this may be a broad absorption feature superimposed on a broad emission feature. We group V1685 Cyg
as a single--peaked emission object based on the emission near zero velocity. The binned spectrum of
T Ori also shows signs of being double--peaked but the spectrum is too noisy be certain. 

\subsection{Double--peaked emission}

All of the double--peaked (DP) emission profiles (6 of 48 objects, or 8\%) show broad, weak emission
extending to velocities of $\pm$200--300 km s$^{-1}$. The velocities of the two emission peaks are
generally not symmetric about zero velocity, with the center of the peak separation having an RV
ranging from $\sim$-60 to 30 km s$^{-1}$. These shifts may be due to real system velocities caused
by massive companions. HD 37490 is included in this group since we are not able to definitively rule
out its classification as a HAEBE. We choose to classify HD 35187 as a DP object due to the fact
that the absorption is centered and the emission peaks are at roughly the same velocities
($\sim$$\pm$200 km s$^{-1}$). The emission may well be masking a mass flow signature but it is not
clear from the observed profile.  

The most interesting thing to note about the DP profiles is the roughly equal equivalent widths of
both the blue and red emission components, with a maximum deviation of $\sim$8\% from a ratio of
unity.  With the exception of HD 35187, the DP profiles seen here do not show any sub--continuum
absorption features; the peaks are always roughly symmetric and of comparable width and strength.
Double--peaked profiles of this nature are often seen in molecular line observations of rotating
disks \citep[e.g.][]{mannings97a,dent05} and are common at H$\alpha$ in HAEBES with evidence of
circumstellar material \citep{viera03}. Double--peaked profiles in Ca II have also been observed in
a number of HAEBES \citep{hp92}. 

Our DP profiles are, for the most part, consistent with the interpretation of Keplerian rotation in
a disk and not with formation in a wind. As the inclination angle of the system decreases, the peaks
will move closer to zero velocity until, at a face--on inclination, the peaks will completely merge
and form a single--peaked emission line. If the emission is being formed in a rotating disk, the
typical velocities seen here indicate an emission region within a few stellar radii of the surface.
For example, HD 34700, which has a maximum blue emission velocity of 175 km s$^{-1}$, has a mass of
2.4 \Msun, a radius of 4.2 \Rsun, and an inclination angle of $\sim25^\circ$ which places the
minimum distance of emission at $\sim$0.25 $R_*$ from the stellar surface. For the hottest objects,
it is possible that the photoionization rate is too high within a few stellar radii for \hei to
form, but a full treatment of the He ionization states and level populations near the star are
beyond the scope of this paper. The lack of detected red--shifted absorption in the DP stars
suggests magnetically controlled accretion is not the mechanism responsible for the emission.
Instead, it is more likely that the DP profiles are formed as the result of Keplerian rotation very
near the stellar surface. In addition, the profile shapes are qualitatively similar to the \hei
double--peaked profiles of classical Be objects which contain disks of ejected material at small
distances from the stellar surface \citep{groh07}. If the disk extends close to the stellar surface,
material will be deposited onto the star via an equatorial boundary layer in which luminosity is
generated by the difference in kinetic energy between the disk gas and the stellar photosphere. The
luminosity generated by the boundary layer--star interaction is sufficient to produce to the excess
Balmer emission seen in some HAEBES \citep{blondel}. In the case of our DP sample, accretion rates
are estimated in the literature for 3 of 4 objects suggesting that accretion is indeed occurring.    

\subsection{Exceptions: MWC 137 and V374 Cep}

MWC 137 and V374 Cep both display unique morphologies. MWC 137 shows very strong, broad emission
with the wings of the profile extending to $\pm$300 km s$^{-1}$. There is a strong absorption
feature superimposed on the emission profile indicating a significant amount of material near zero
velocity. We note that the RV for this object is unknown due to the lack of observable photspheric
features. If the extended wings of the profile are actually centered at $\sim$0 km s$^{-1}$, the
absorption feature would be blue--shifted to $\sim$-20 km s$^{-1}$. This narrow absorption is the
only narrow blue absorption observed in our sample; however, this absorption is broader than
expected for interstellar absorption. A reliable estimate of MWC 137's RV will provide a clearer
picture of the nature of the circumstellar environment.

V374 Cep shows a very strong absorption feature near zero velocity that is superimposed on a broad
emission profile extending to $\sim$$\pm$300 km s$^{-1}$. This is the strongest absorption feature
in our sample. Similar absorption features are present in the balmer lines. This feature is
comparable to those of UX and GK Tau at \hei from \citetalias{EFHK}. No information exists on the
existence of a disk around V374 Cep. If a disk is present, it is possible we are looking at the
system close to edge--on and the deep absorption is a result of the optically thick disk absorbing
any stellar or accretion emission. The broad emission component could be the result of an extended
wind surrounding the system. We are confident that this is not a strong photospheric absorption
feature due to the lack of any photospheric absorption lines in B and A--type stars at 10830
\AA\hspace{0pt} with line depths greater than $\sim$0.3. The absorption seen here penetrates to a
depth of $\sim$0.75, ruling out a stellar origin.    

\section{DISCUSSION}

Our sample displays a wide variety of line profile morphologies. The incidence of red and
blue--shifted absorption in HBe stars is significantly different than in CTTSs. For HAe stars, only
the incidence of blue--shifted absorption differs significantly from that in CTTSs; red--shifted
absorption appears in similar fractions of both groups of objects. Taken as a single group, both the
red and blue--shifted incidence in HAEBES differs significantly from that in CTTSs. Below we
elaborate on the implications of our data concerning the accretion and outflow geometry around
HAEBES. We also discuss how the data highlight apparent differences between Herbig Ae and Be stars
and between HAEBES and CTTSs.

\subsection{Interpretation of line profile morphologies}

In this section we will focus on the red and blue--shifted absorption profiles and interpret their
key features within the context of accreting and outflowing material.

\subsubsection{Winds: P--Cygni and blue--shifted absorption} 

In the discussion below of the observed blue--shifted absorption profiles, we largely base our
interpretations on the models of \citet[][herafter KEF]{KEF}\defcitealias{KEF}{KEF}.
Although these models are calculated using typical CTTS parameters, the general trends in
profile morphology with inclination angle and inner disk radius should also hold for HAEBE
parameters. \citetalias{KEF} model the \hei profiles of CTTSs by assuming that the absorbing atoms
are either (1) emerging radially away the central star in a spherically symmetric \textit{stellar
wind}, (2) emerging radially from the star at angles less than 60$^{\circ}$ from the pole in a
\textit{polar} stellar wind, or (3) flowing away from the disk along streamlines at a fixed angle of
45$^{\circ}$ from disk--normal in a \textit{disk wind}. The polar stellar wind is meant to emulate
the MA scenario in which accreting material is predominantly deposited at high latitudes, i.e. an
accretion powered stellar wind \citep{matt05}. The deposition of energy mainly at high latitudes is
believed to drive a stellar wind more strongly from these locations. In a spherically expanding
stellar wind, there is no preferred location on the stellar surface from which the wind is driven.
It is unclear how MA can produce a wind of this nature. Thus the polar wind provides an
approximation for what a stellar wind might look like if driven by high--latitude MA flows. Their
models include the scattering of stellar continuum photons and, in some cases, an in situ emission
component. The stellar wind models also follow multiple scatterings of photons back into the line of
sight. They explore a range of disk truncation radii (resulting in disk shadowing of the retreating
flow, in some cases), disk inclination angles, and high and low turbulent gas velocities ($V_H$ and
$V_L$), all of which can have a significant impact on the resulting profile morphologies.
Disk--shadowing is only important for the stellar wind calculations. Many of our \hei profiles
qualitatively match the computed model profiles of \citetalias{KEF}, readily facilitating
interpretations of the outflow geometries producing our observations. In addition, \citetalias{KEF}
use a featureless continuum in their models. The HAEBES in our sample generally show no photospheric
features at \hei ensuring that comparisons to \citetalias{KEF}'s profiles will only be made between
circumstellar features in both cases. Rotation, which could be a complicating factor due to
the large $v$sin$i$ values of many objects in our sample, is not included in KEF's stellar wind
models. We note that \citetalias{KEF} do not calculate line formation in the actual accretion
flow.

In general, our observed profiles match better the stellar wind profiles of \citetalias{KEF}, and
those are the ones we discuss further. \citetalias{KEF} show that the emission peaks near zero
velocity can be reproduced if the receding flow is blocked by the disk, i.e. the disk either extends
to the surface of the star or the truncation radius is well inside the radius of origin of the
outflow (Figures 4 and 5 of \citetalias{KEF}). The PCII type profiles in our sample are well matched
by the disk--shadowing scenario of \citetalias{KEF} which produce centered emission of comparable
strength to the blue--shifted absorption. It is not possible to be more specific concerning the
inclination angle and truncation radius of the system based on our observed profiles using only the
qualitative similarities of different models from \citetalias{KEF}. However, since the disk
shadowing scenario is plausible and since it is difficult to produce our PCII profiles without some
degree of disk shadowing, it is reasonable to conclude that this geometry applies to the PCII
objects. We note that the polar stellar wind models, which is the likely case for accretion powered
stellar winds where material is impacting the stellar surface along magnetic field lines at high
latitudes, tend to produce stronger emission than absorption for non--polar viewing angles. The
polar wind profiles do not match our PCII profiles as well as the spherical wind model profiles.

The PCI profiles tend to be matched better by models in which disk shadowing is unimportant, i.e.
the disk truncation radius, $R_t$, is $\sim$2--5 $R_*$ allowing the red--shifted emission peak to
become visible. However, this type of profile is also produced if the inclination angle is small for
a disk--shadowed system, allowing the red--shifted emission to be viewed through the small space
between the star and truncated disk. More importantly, the PCI profiles can also be produced by disk
shadowed or non--disk shadowed systems that are viewed nearly edge--on allowing much of the receding
flow to be visible above and below the disk. In both the disk--shadowed and non--disk--shadowed
scenarios it is possible to produce the red--shifted peaks for a wide range of viewing angles,
though a high turbulent gas velocity is required in all cases. On the other hand, the lack of
detected magnetic fields on HAEBES makes large $R_t$, and thus the absence of disk--shadowing,
unlikely. The polar wind profiles of \citetalias{KEF} do not show red--shifted peaks and thus cannot
account for the PCI profile morphologies. The profiles of AB Aur, HD 163296, HD 38120, and HD 250550
all have comparable emission and absorption components implying that some form of in situ emission
is necessary to balance the profile contributions. Objects with broad, deep absorption and weaker
emission such as MWC 758, MWC 1080, and MWC 480 are better matched by profiles which only include
scattering and disk shadowing is unimportant. It appears that a variety of scenarios are potentially
responsible for the formation of the PCI profiles. However, due to the evidence for smaller
magnetospheres, and thus smaller disk truncation radii, presented in the next section we believe it
is less likely that the PCI profiles are produced in systems with large $R_t$.  

The profile of V380 Ori is unique among our PC objects: the emission is centered at zero--velocity
and strongly dominates the blue--shifted absorption. This type of profile is most commonly produced
by the polar wind models of \citetalias{KEF} in which the viewing angle is $\sim$60$^\circ$ and in
situ emission is present. V380 Ori is a confirmed magnetic HAEBE \citep{wade05,alecian13} and
appears to be strongly accreting \citep{db11}. In addition, it shows clear signs of significant
circumstellar material \citep{hillenbrand92}. Thus it is possible that the polar wind signature is
the result of a stellar wind driven by magnetically controlled accretion onto the star at high
latitudes. However, if the viewing angle is $\sim$60$^\circ$ and the accretion stream is impacting
the star at $<$60$^\circ$ from the pole, then red--shifted absorption should appear in the
profile. This scenario will be explored more fully using optical accretion diagnostics (Cauley \&
Johns--Krull 2014, in prep).

\subsubsection{Accretion: Inverse P--Cygni and red--shifted absorption}

In CTTSs IPC signatures are produced by MA \citep[e.g.][]{edwards94,alencar00}. As was previously
mentioned in \S 2, the lack of magnetic fields detected around HAEBES casts doubt, but does not rule
out the possibility, that the MA scenario holds for most of these objects. This is especially true
for the HBe stars which show a much lower incidence of red--shifted absorption at \hei than CTTSs
(\autoref{tab:tab4} and \autoref{tab:tab6}). None of the objects displaying IPC profiles in our
sample have detected magnetic fields \citep{alecian13}.
\citet{fischer08}\defcitealias{fischer08}{FKEH}, hereafter FKEH, modeled the red absorption at \hei
in CTTSs by assuming that stellar and accretion--generated continuum photons are scattered by a
funnel flow falling along a dipolar magnetic field geometry from an annulus in an accretion disk.
Their models show that for all but the smallest inclination angles, i.e. a viewing angle nearly
pole--on, red--shifted absorption is prominent for almost the entire range of funnel flow
parameters. \citetalias{EFHK} observe red--shifted sub--continuum absorption in at least one
observation in 19 of 39 CTTSs, 49\% of their sample, all of which are known to be accreting. Even
excluding the objects that have variable \hei profiles and do not show red absorption in all
observations, 18 of 39, or 46\%, of their objects show red--shifted absorption.  Thus the detected
36\% of HAe stars and 15\% of HBe stars with IPC profiles in our study (\autoref{tab:tab4}) suggests
that MA is less common among the HBe stars than it is among CTTSs and occurs at a similar rate in
the HAe stars compared to CTTSs. This is confirmed by contingency tests performed on the different
groups (\autoref{tab:tab6}).  

\begin{figure}\label{fig:fig5}
  \begin{center}
     \includegraphics[scale=.58,clip,trim=30mm 0mm 40mm 0mm]{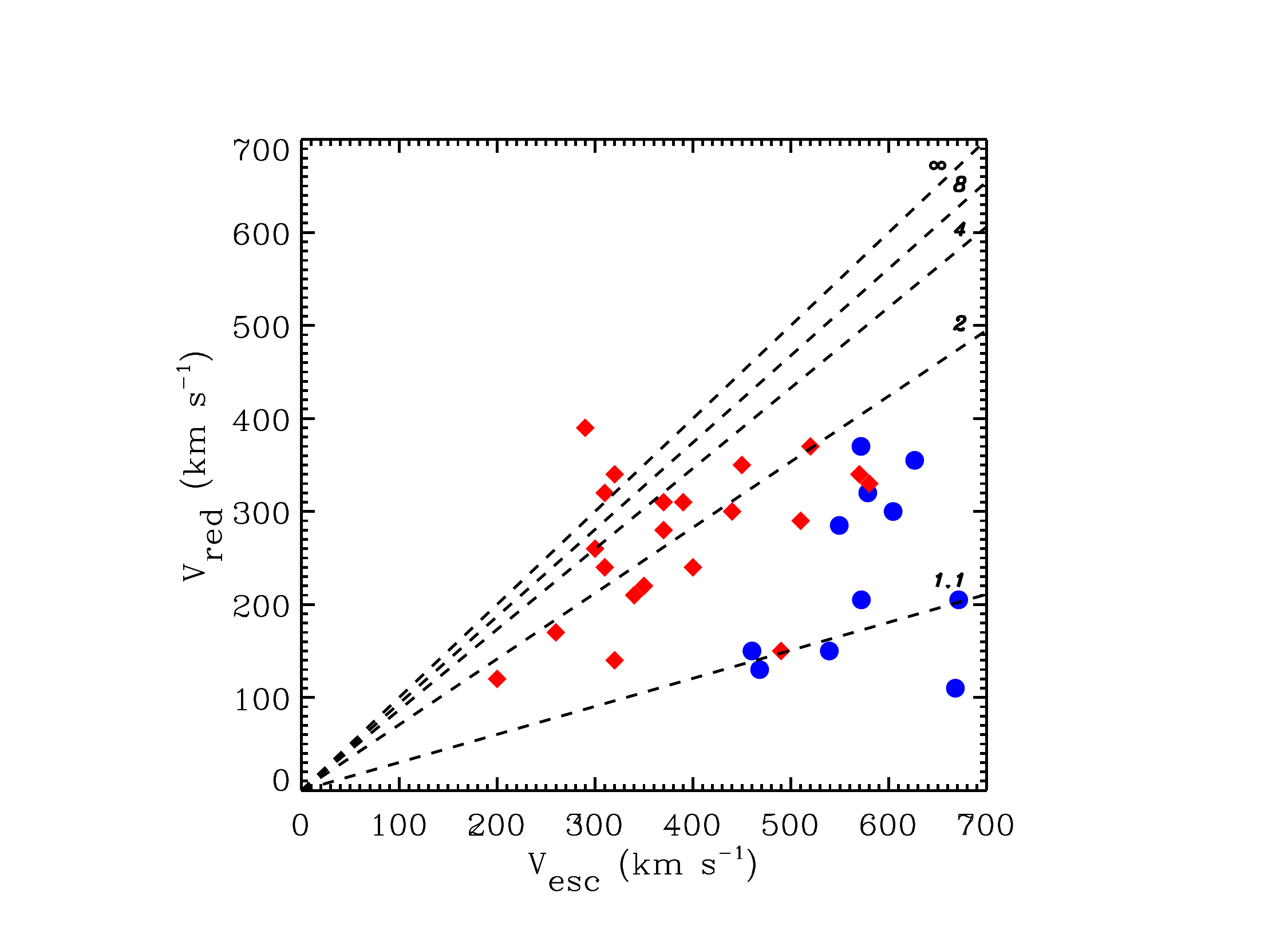}
     \caption{The maximum red--shifted absorption velocity versus escape velocity for the
red--shifted absorption HAEBES in our sample (blue circles) and the
CTTSs from \citetalias{fischer08} (red diamonds). This plot is a reproduction of Figure 5 from
\citetalias{fischer08} with the addition of the red--shifted absorption HAEBES from this study. The
diagonal dashed lines indicate the maximum infall velocity attained by material originating from the
distance (given in units of $R_*$) indicated on the right hand side of the plot. The HAEBES on
average display maximum absorption velocities that are a smaller percentage of their escape
velocities suggesting that material is accreting onto the star from closer to the stellar surface.}
  \end{center}
\end{figure}

If HAEBES experiencing MA have smaller magnetospheres, the maximum infall velocities reached by the
accreting gas should be smaller fractions of the stellar escape velocity than is found for CTTSs.
\citetalias{fischer08} measured the approximate maximum red--shifted absorption velocity for the
CTTS \hei profiles from \citetalias{EFHK}. We have reproduced \autoref{fig:fig5} from
\citetalias{fischer08} with the inclusion of the red--shifted absorption HAEBES from this study.
\autoref{fig:fig5} shows the measured maximum red--shifted absorption velocity plotted against the stellar
escape velocity for the IPC and red--shifted absorption HAEBES in our sample (blue circles) and the
CTTSs from \citetalias{fischer08} (red diamonds). It is clear that CTTSs on average show maximum
red--shifted absorption velocities that are a greater fraction of their stellar surface escape
velocities. This suggests that accretion flows in HAEBES begin deeper in the system's gravitational
potential, i.e.  closer to the star, as can be readily understood by noting the relationship between
$V_{red}$ and $R_{red}$, the distance at which the infalling material originates, given in equation
2 of \citetalias{fischer08}. This is what one would expect if the magnetospheres, and thus the disk
truncation radii, are smaller in HAEBES than in CTTSs.

\begin{figure}\label{fig:fig6} 
  \begin{center} 
    \includegraphics[scale=.55,clip,trim=20mm 15mm 0mm 0mm]{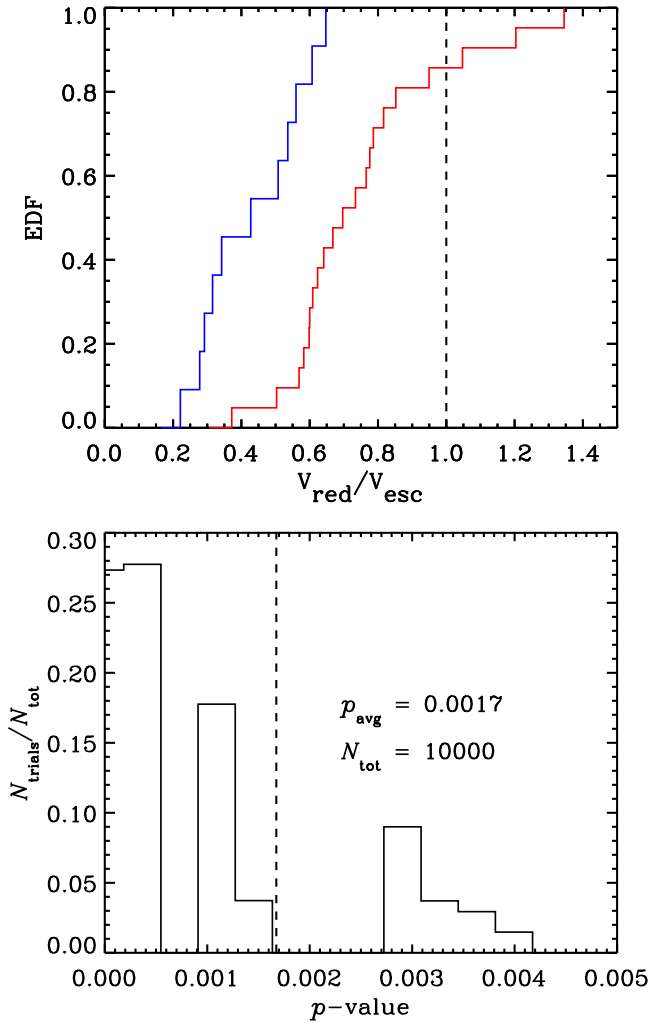} 
    \caption{The emipirical distribution functions (top panel) for the HAEBES (blue line) and 
CTTSs (red line) from \autoref{fig:fig5} and a histogram of two sample KS test $p$-values (bottom panel) for 10,000 random
samplings, limited by an assigned 28\% uncertainty in the stellar escape velocity, of the observed
velocity ratio distributions. The vast majority of $p$--values are $<$0.01 indicating that the
distributions are different at a confidence level $>$99\%.} 
  \end{center} 
\end{figure}

In order to test the significance of the differences between the HAEBES and CTTSs in
\autoref{fig:fig5}, we have performed two--sample KS tests on ten thousand random samplings of the
observed distributions of the ratio of red--shifted absorption velocity to stellar escape velocity.
In order to specify a range of allowed values for each sampling, we have assigned a 50\% uncertainty
in the stellar mass and a 25\% uncertainty in the stellar radius to each object which allows the
escape velocity to vary by 28\%.  We believe these values to be reasonable based on typical
published uncertainties \citep[e.g.][]{alecian13}. Each sampling is assigned a random escape
velocity within the computed 28\% range. No uncertainty is assigned to the measured red--shifted
velocity. The empirical distribution function for each group is shown in the top panel of
\autoref{fig:fig6}; a histogram of the $p$--values from the KS tests is shown in the bottom panel.
It is clear from the distribution of $p$--values, where greater than 99\% of the resulting values
are less than 0.01, that there is a significant difference between the ratio of red--shifted
absorption velocities to stellar escape velocities for HAEBES and CTTSs. This result confirms that
the distributions are different at the $>$99\% confidence level and supports the hypothesis that the
magnetospheres in HAEBES experiencing magnetospheric accretion are smaller than in CTTSs.

\subsection{Outflows and Accretion}

\subsubsection{Magnetospheric and boundary layer accretion}

Magnetospheric accretion appears to be uncommon in mid to early B--type stars but seems to be
present at a similar rate to CTTSs in HAe and late HBe stars. Useful information concerning the
accretion geometries of the stars in our sample can be gained from narrowing our analysis to only
the stars with determined accretion rates. If HAEBES occupy a broad range of evolutionary states, it
is certainly possible that many negligibly accreting objects are present in our sample which would
lower the occurrence rate of IPC profiles at He I $\lambda$10830, although we have attempted to
remove these objects from the sample (see \S 3.1). \citetalias{EFHK} and \citetalias{fischer08} show
that red--shifted absorption signatures at \hei are absent for CTTSs with 1--$\mu$m veilings $>$0.5,
with veiling defined as the ratio of the excess emission to the underlying stellar photosphere
\citep[see][]{hartigan89}. Higher veiling at 1 $\mu$m has been shown to scale with other accretion
indicators in CTTSs, indicating that objects with higher accretion rates will show higher veilings
\citep{fischer11}. \citetalias{fischer08} suggests this may be the result of different accretion
geometries becoming important in CTTSs with higher accretion rates, i.e. MA may be inhibited by
magnetospheres being crushed down to the stellar surface due to high disk accretion rates. Smaller
magnetospheres in HAEBES could mimic this result since weaker red--shifted absorption tends to be
produced by these geometries compared to larger magnetospheres \citepalias{fischer08}. These weak
signatures may be masked by emission in an outflow and further reducing the incidence of red-shifted
absorption.

\begin{figure}\label{fig:fig7}
  \begin{center}
     \includegraphics[scale=.55,clip,trim=22mm 0mm 0mm 10mm]{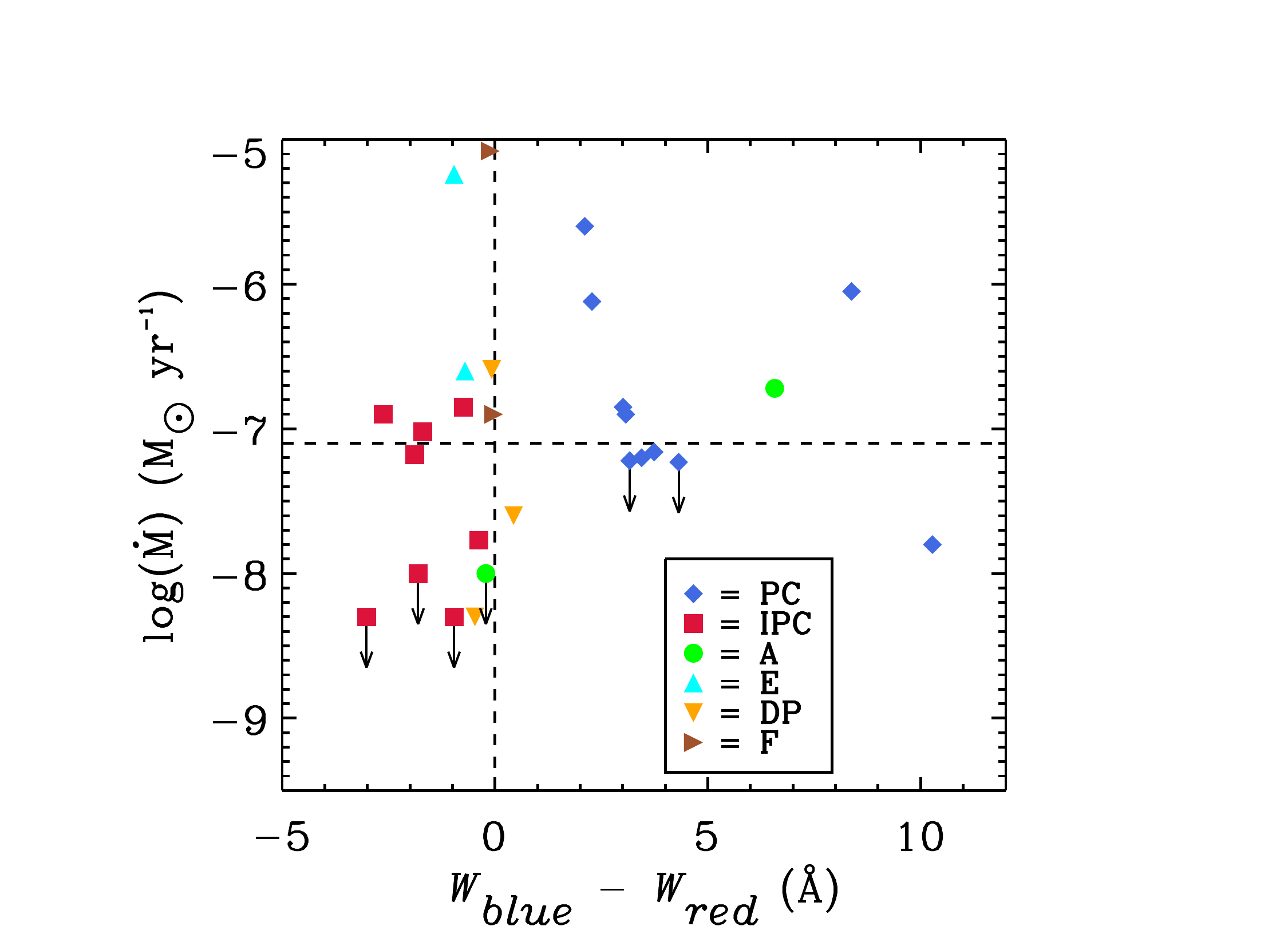}
     \caption{Blue equivalent widths minus red equivalent widths versus accretion rate for the stars
in our sample with measured accretion rates. The profile type abbreviations in the legend are the
same as those used in \autoref{fig:fig4} and \autoref{tab:tab3}. Upper limits on accretion rates are marked with a
downward arrow. The horizontal dashed line is the mean accretion rate for the objects in our sample.
There is a dearth of stars with P--Cygni--like profiles at low accretion rates and a relative lack
of red--shifted absorption objects with large accretion rates.}
  \end{center}
\end{figure}

To test this hypothesis in our HAEBE sample, in \autoref{fig:fig7} we have plotted the mass
accretion rate (column 7 from \autoref{tab:tab2}) versus the blue EW minus the red EW. The EW difference is
essentially a measure of profile type since negative values imply IPC profiles and positive values
imply P--Cygni profiles; objects with centered emission or absorption, or no features, should be
near zero.  Accretion rates with upper limits are indicated by the downward arrows. The mean
accretion rate for our sample (log($\dot{M}$)\ltsima -7.1) is indicated by the horizontal dashed
line.  \autoref{fig:fig7} clearly shows that there is a relative lack of objects with low accretion
rates which display P--Cygni profiles. In other words, the objects in our sample with the lowest
accretion rates predominantly show red absorption in their \hei profiles and the highest accretion
rate objects are PC objects.  There are also four objects with average accretion rates that show red
absorption.  Thus, the trend seen in CTTSs for highly accreting stars to not show red--shifted
absorption seems to be present in our HAEBE sample. We note that there is no significant statistical
difference between the accretion rate distributions of the PC and IPC objects in \autoref{fig:fig7},
i.e. a KS test results in a $p$--value of 0.38. 

Our spectra are not simultaneous with the observations used to determine the values of $\dot{M}$ and
accretion onto young stars is intrinsically variable \citep{bouvier03}.  Accretion
\textit{signatures} can also be variable due to the combined geometric effects of stellar rotation
and magnetic field inclination angle, mimicking real changes in the accretion rate
\citep[e.g.][]{symi05,bouvier07a}. A more complete data set would combine simultaneous, independent
measurements of both $\dot{M}$ and \hei EW. \citet{mend11} estimate that HAEBE accretion rates in
general vary by $\sim$0.5 dex, or a factor of $\sim$3, on timescales of days to months. The He I
$\lambda$5876 line, which immediately precedes \hei in a recombination--cascade sequence, has a
highly variable morphology in some objects \citep{mend11a}. However, the total He I $\lambda$5876
equivalent width only varies by $\sim$1.5 \AA in even the most highly variable objects. Since we
might expect changes in He I $\lambda$5876 to reflect similar changes in He I $\lambda$10830, even
the largest changes in equivalent width would only impact the 3 red--shifted absorption objects in
\autoref{fig:fig7} with equivalent width differences of $<$-1.5. Thus while variability may
certainly impact individual objects, we believe the general trend seen in \autoref{fig:fig7} will
not be significantly altered.  This should be tested with simultaneous determinations of the
equivalent width and accretion rate, as mentioned above. 

If red--shifted absorption at \hei in CTTSs with large accretion rates tends to be suppressed due to
a crushed magnetosphere, then the relative lack of red--shifted absorption in HAEBES with large
accretion rates suggests that a similar scenario is operating around these objects. The weaker
magnetic field strengths on HAEBES compared to CTTSs supports the idea that large accretion rates
are able to force the magnetosphere down to smaller radii, with the extreme cases (e.g. very weak
magnetic fields or very large accretion rates) potentially resulting in boundary layer accretion. If
the magnetospheres around HAEBES are \textit{intrinsically} smaller and the field strengths are
weaker, then increases in the accretion rate may more easily crush the magnetosphere closer to the
stellar surface, resulting in the lower incidence of red--shifted absorption in HAEBES with large
accretion rates. \autoref{fig:fig7} also shows a relative lack of blue--shifted absorption at low
accretion rates. If HAEBE magnetospheres are crushed by large mass accretion rates, this would
indicate that the outflows around HAEBES are predominantly driven by boundary layer--type accretion
since the blue--shifted absorption profiles are predominantly produced by objects with large
$\dot{M}$.  However, as pointed out by \citetalias{fischer08} for the CTTSs, the lack of
red--shifted absorption could also be due to the filling in of red--shifted absorption by wind
emission and not the altering of the accretion geometry. The lack of \textit{simultaneous} red and
blue--shifted absorption in our sample, however, suggests it is the latter since simultaneous
accretion and outflows would likely result in observable simultaneous blue and red--shifted
absorption in at least some objects.

If many accreting HAEBES have smaller magnetospheres or are accreting through a boundary
layer, then the truncation radius will be smaller or nonexistent. The lack of a truncated disk has
implications for planet migration down to small orbital radii. It was first suggested by
\citet{lin96} that planetary migration may eventually be halted by a resonance interaction with the
truncated inner--disk. If this mechanism contributes to halting inward planet migration, and
sufficiently truncated disks are not present in many HAEBE systems, then the frequency of gas giant
planets in close orbits ($r$$<$0.10 AU) should be much lower than for solar--mass objects. We note
that simulated density functions by \citet{plavchan13} recently found evidence that this
migration--halting scenario does not match the orbital distribution of Jupiter--mass planets as well
as a tidal circularization model does.  However, this analysis was performed for objects with masses
$<$1.5 $\Msun$ due to the small statistics for planets around F, A, and B--type stars. A larger
sample of orbital radii for giant planets around F, A, and B stars will provide a better basis for
comparing how the lack of inner disks during early stellar evolution can affect planet migration.

\subsubsection{Disk winds}

When comparing the \hei profiles of our HAEBES to the CTTS profiles in \citetalias{EFHK}, a feature
that differs significantly between the two samples is that \textit{the HAEBE sample shows no
evidence of narrow, blue--shifted absorption at any velocity}. We exclude the possible case of MWC
137 due to its unknown system RV. According to the models of \citetalias{KEF} the key profile
signature of a disk wind\footnote{In this context we are referring to disk winds formed near the
star--disk interaction region and not to disk winds formed at large radial distances from the star.
Temperatures and photo--exciting radiation fluxes are too low at large distances in the disk to
produce any significant excitation of \hei.} is narrow blue--shifted absorption resulting from the
scattering of stellar photons along streamlines extending outward at some angle from the disk. The
small range in absorbed velocities is due to the fact that the line of sight to the stellar surface
only intersects a small range of velocities in the disk wind, compared to the entire range of
velocities, including the acceleration zone, for wind streamlines that originate on the star. As the
inclination of the system becomes more edge--on, the absorption tends to become stronger.

\citetalias{KEF} find $\sim$29\% of the \citetalias{EFHK} CTTS sample to be good disk wind
candidates at He I $\lambda$10830; our sample contains no strong disk wind candidates. Outflows
launched from near the interaction region of the disk and stellar magnetosphere are a natural
consequence of magnetically controlled accretion \citep[e.g.][]{mohanty08,kurosawa12,zanni13}. Thus
it is expected that some type of disk wind signature would be present in a non--zero fraction of
accreting HAEBE profiles if MA involving large magnetospheres were the dominant accretion mechanism.
On the other hand, if HAEBES tend to have smaller magnetospheres than CTTSs, a disk--wind launched
from a range of radii near the truncation point closer to the stellar surface would become
indistinguishable from a stellar wind for most viewing angles. This is especially true for X--wind
configurations where the wind launching region is confined to a very narrow annulus near the
corotation radius \citep{shu94}. Thus we cannot rule out the presence of disk winds in our sample
based on the lack of narrow blue--shifted absorption profiles. We note that the lack of
direct inner disk wind evidence in our sample does not conflict with the recent interpretations of
interferometric observations of Br$\gamma$ in a small number of HAEBES as arising in a disk wind
\citep{tatulli07,kraus08}. The disk winds diagnosed in these cases \citep[specifically DX Cha, HD
163296, and V921 Sco][]{kraus08} likely arise over much larger radii (i.e. near the dust sublimation
radius at a few tenths of an AU) than are traced by He I $\lambda$10830.     

\subsubsection{Accretion--driven outflows in the IPC objects}

There is an important difference between the IPC profiles in our HAEBE sample and the CTTSs from
\citetalias{EFHK}: blue--shifted absorption, indicating a disk or stellar wind, is present in 16 of
21 CTTSs that also show red--shifted absorption. The 13 stars with red--shifted absorption or IPC
profiles in our sample \textit{do not show any evidence of blue--shifted wind signatures}. There is
a strong correlation between accretion and outflow rates in TTSs \citep[e.g.][]{hartigan95} which
points to mass accretion being the main driver of outflows in these objects. The presence of outflow
signatures in CTTS \hei IPC profiles also supports this conclusion. The relationship between mass
accretion and mass outflow extends into the HAEBE mass range \citep[e.g.][]{ellerbroek13}.
Since this relationship seems to apply to HAEBES, we would expect to see some sort of outflow
signature at \hei in a fraction of the IPC objects. Since this is not observed, and since literature
H$\alpha$ profiles \citep[e.g.][]{fink84,hp92,mend11a} of the IPC objects do not show any direct
evidence of blue--shifted wind signatures, it is currently unclear whether or not mass accretion is
capable of driving strong outflows on these objects. As noted in \S 5.2.1 the IPC objects generally
have average or below--average accretion rates. Based on the $\dot{M}_W$ vs $\dot{M}_{acc}$
relationship from \citet{ellerbroek13}, accretion rates of this magnitude should be able to drive
detectable outflows. Similar average accretion rates ($\sim$10$^{-7}$ $\Msun$ yr$^{-1}$) are seen in
some of the PC objects confirming that these accretion rates are capable of driving outflows. This
suggests that the IPC objects, which show evidence of MA, are less efficient at driving outflows
compared to the PC objects which may be accreting through a different mechanism (see \S 5.2.4). 

The presence of red--shifted absorption at \hei requires the existence of infalling material near
the star. The lack of outflow signatures in our IPC objects suggests that these stars do not have
strong accretion--powered outflows. It is unlikely that this is due to differing formation
conditions for the wind versus the accretion flow since \hei signatures tracing outflowing gas is
expected to form at similar distances from the star as the accreting material. In addition, the
relative lack of strong outflows in HAEBES as a whole is supported by optical forbidden line
statistics which show a much lower fraction of occurrence of large scale outflows in HAEBES than in
CTTS \citep[e.g.][]{hp92}. If HAEBE magnetospheres are smaller than those of CTTSs, it may be more
difficult to produce the magnetic configurations at the disk--magnetosphere interaction region
necessary to drive strong outflows. Furthermore, mass infall from deeper in the star's gravitational
potential will result in less of the necessary energy to accelerate an outflow from the stellar
surface. Thus accretion driven \textit{stellar} winds in HAEBES experiencing MA may be less
efficient. To our knowledge, no simulations of the magnetospheric star--disk interaction region have
been performed using typical HAEBE parameters. MHD calculations using HAEBE parameters would be
informative.      

\subsubsection{Accretion--driven outflows in the PC objects}

None of the P--Cygni objects in our sample show evidence of mass infall in He I $\lambda$10830.
\citetalias{fischer08} showed that weak red--shifted absorption is produced at most inclinations
angles for stars with low veilings and truncation radii of $\sim$2 $R_*$. These weak absorption
signatures could be overwhelmed by strong emission in a wind \citep{KEF}. Red--shifted absorption
from a MA flow would also be absent for inclination angles near $\sim$0$^{\circ}$. Thus it is
possible to see evidence of an accretion--launched outflow without seeing any direct signature of
mass infall. However, the significant lack of red--shifted absorption seen in HBe \hei profiles
compared to HAe stars and CTTSs suggests mass infall is truly absent rather than being the result of
viewing angle effects or filling in by wind emission. An alternative to MA is to drive the outflows
from an equatorial boundary layer created by the accretion disk at the stellar surface.  This will
be discussed below in \S 5.4. We note that only one object in our sample has an estimated luminosity
high enough (MWC 1080; $L_*$$\sim$3.8 x 10$^4$ $\Lsun$) to potentially produce a radiatively driven
wind of any significance \citep{krticka}. Thus the strong outflows observed in our sample are most
likely accretion related.   
 
\subsection{Herbig Ae vs Be stars}

There is growing consensus that Herbig Ae (HAe) stars interact with their environments differently
than Herbig Be (HBe) stars. In particular, HBe objects are found to show weaker evidence for MA,
including line widths inconsistent with gas rotating at Keplerian velocities near or inside of the
disk truncation radius \citep{mend11}, a lower incidence of line polarization effects due to
scattering of stellar photons by disk--like circumstellar material \citep{vink02,mott07},
differences in excitation conditions of H$_2$ \citep{martinz08}, and a possible breakdown of the Br
$\gamma$--accretion luminosity relationship that holds for CTTSs and HAe stars \citep{db11},
although we note that the accretion luminosities determined by \citet{db11} are calculated
with a relationship derived using an A2V photosphere which may under--predict $L_{acc}$ for stars
with spectral types earlier than A2. A preliminary analysis of \hei data presented by \citet{oud11}
for 79 HAEBES shows a shift from absorption dominated profiles in HAe objects to more emission
dominated profiles in HBe objects \citep[Figure 3 from][]{oud11}. We note that our data do not
display this trend: we find no relationship between the total \hei EW and effective temperature. The
observed differences between HBe and HAe objects may indicate that, in general, the two groups are
being observed at different phases of their pre--main sequence evolution. More specific evolutionary
classifications, as proposed, for example, by \citet{malfait98}, would facilitate comparisons
between objects that are at more similar stages of development. A second, although related, possible
source of the observed differences is the misclassification of \textit{classical} Be objects as HBe
stars (see \S 4.1). It is unlikely that the PC and IPC HBe stars in our sample are misclassified
since their H$\alpha$ features, in general, do not match the typical zero--velocity centered
double--peaked emission of the classical Be objects \citep{hanu96}.

\begin{figure}\label{fig:fig8}
  \begin{center}
     \includegraphics[scale=.58,clip,trim=25mm 0mm 0mm 13mm]{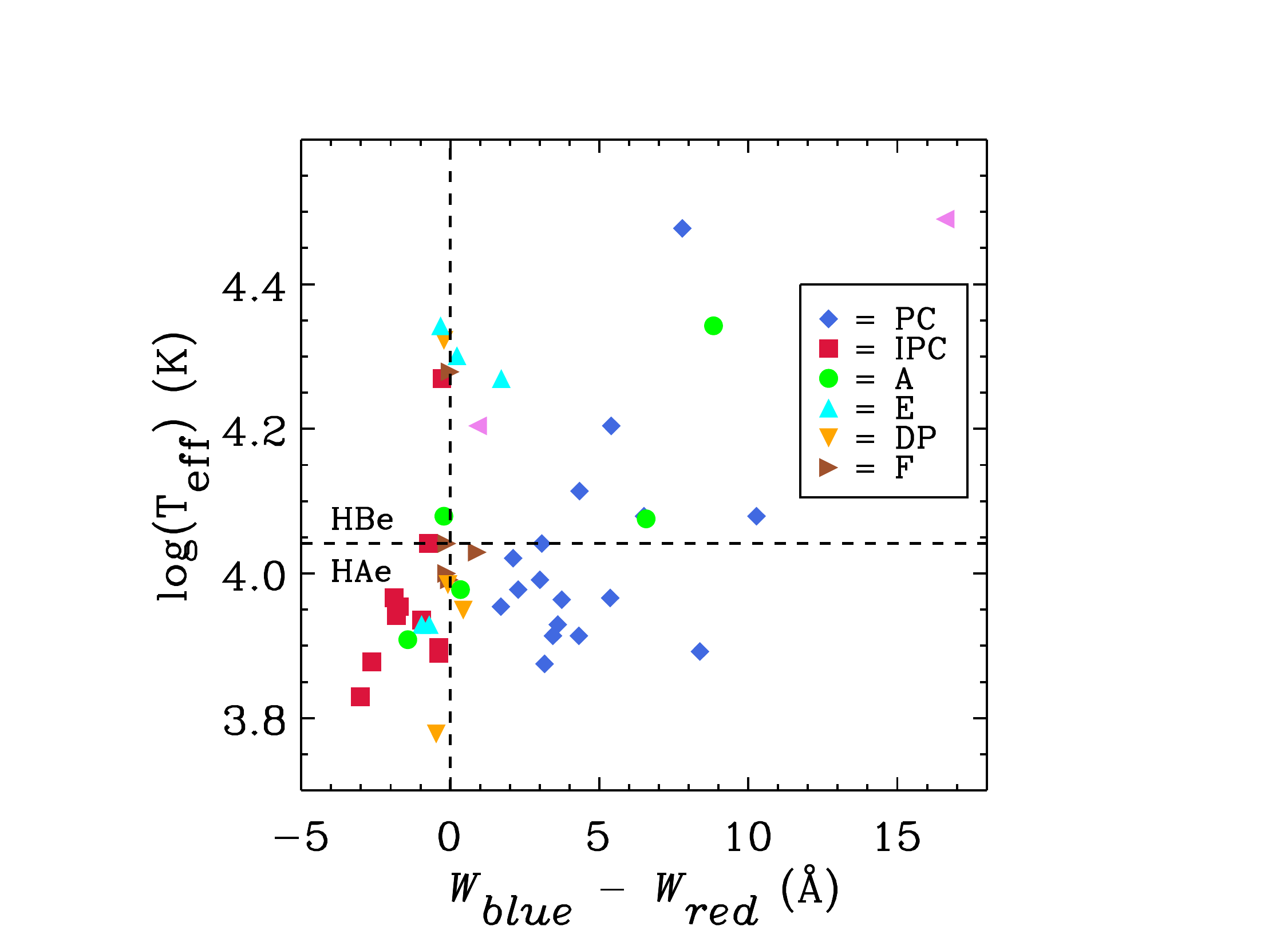}
     \caption{Effective temperature versus blue minus red \hei equivalent width. Plotting symbols
and colors are the same as in Fig. 7 with the exception of V374 Cep and MWC 137 which are plotted as a
left--pointing purple triangles. An approximate separation between A and B spectral types is
marked by the horizontal dotted line. Potentially misclassified objects from \S 3.1 are excluded. 
Note the lack of B--type stars in the upper left quadrant.}
  \end{center}
\end{figure}

Assuming that all of the objects in our sample not discussed in \S 4.1 are actually HAEBES, the HBe
stars show different rates of occurrence of particular line profile morphologies compared to the HAe
stars (\autoref{tab:tab4}), although the contingency tests result in relatively high $p$--values
(\autoref{tab:tab6}) indicating that the red and blue--shifted absorption incidence is not
significantly different between the two groups. The statistics are illustrated in \autoref{fig:fig8}
where it is shown that any significant red--shifted absorption is almost completely absent from the
HBe sample (symbols above the horizontal dotted line). There is only one early B--type object (HD
200775, a B4 star) that displays a red--shifted absorption profile. Note that MWC 137 is included in
the figure (right-- most left--facing purple triangle), although it is excluded from the statistics
in \autoref{tab:tab4} due to the lack of a reliable RV estimate. The HAe stars show a diverse variety of both
PC and IPC profiles, indicating greater overall levels of accretion and outflow activity. 

Although the differences in line profile morphology are not statistically significant for our
sample, \autoref{fig:fig8} provides a hint that physical differences do exist between the HAe and HBe
environments. Thus we argue that our data add to the recent evidence suggesting that there is a
distinction between HAe and HBe stars concerning the circumstellar environment within a few stellar
radii of the star. The late B--type objects are likely more similar to HAe stars than to early
B--type objects. The greater incidence of IPC profiles in HAe stars suggests that mass accretion in
a substantial fraction of these objects is occurring along magnetic field lines. HBe objects show a
much lower incidence of IPC morphologies, although the sample size is small. Outflows seem to be
present in similar numbers in both populations. The observed differences in the line profile
statistics could be due to differing physical conditions in the immediate circumstellar environments
of HAe and HBe stars.  On the other hand, if the conditions around both groups of objects are
similar, our data would indicate that accretion in HBe objects, and potentially some HAe stars, is
occurring through a different mechanism, e.g. a boundary layer at the stellar surface. This was been
previously suggested by \citet{mend11}. Boundary layer accretion would naturally explain the
relative lack of red--shifted absorption features in HBe stars. The questions concerning the
physical conditions in the immediate circumstellar environment can be further addressed with a
larger sample size using optical and UV data. We are currently pursuing this (Cauley \& Johns--Krull
2014, in prep).

\subsection{Outflows driven from the boundary layer}

If boundary layer accretion is partially responsible for the lack of red--shifted absorption
signatures in HBe stars, and possibly for the smaller observed incidence in HAe stars compared to
CTTSs, a natural question arises: can boundary layer accretion drive outflows in the same way as MA
since outflows are seen in an appreciable fraction of both HAe and HBe stars?  The luminosity
released by the boundary layer differs from the luminosity of a magnetically channeled funnel flow
by a negligible factor, at least for reasonable stellar parameters. Thus the energy required to
drive outflows from near or at the stellar surface is readily available in the boundary layer
paradigm. In current 3D MHD models of accretion--generated outflows around TTSs, magnetic fields
provide the main acceleration mechanism and geometric constraint for the expanding wind
\citep[e.g.][]{roma05,roma09}. As we have pointed out, there is currently no observational support
for the existence of magnetic fields on the majority of HAEBES which are required to drive
and support these flows. However, it has also been suggested that accretion--powered stellar winds
are driven by a combination of thermal and centrifugal effects, the thermal effects being driven by
the accretion--heated stellar corona \citep{matt05}. Alfv\'{e}n waves excited by accreting material
may also play a role in driving the wind, although this mechanism also requires a magnetic field. 

X--ray observations have shown a fraction ($\sim$50\%) of HAEBES to be x--ray emitters, although
this is an upper limit since it is unclear if the x--ray emission is intrinsic to the HAEBES or from
unresolved low--mass companions \citep{stelzer06,stelzer09}. The small detected HAEBE
x--ray--to--bolometric luminosity ratios \citep[log($L_x$/$L_{bol}$)$\sim$-6.3;][]{stelzer09} and
the moderate fraction of x--ray emitters detected suggests that most HAEBES do not have sufficiently
active coronae to drive strong Alfv\'{e}n waves or thermal winds. It is unclear what effect the
deposition of energy near the stellar equator will have compared to the higher latitudes that are
characteristic of magnetically channeled accretion. However, one possible scenario is the
X--celerator mechanism proposed by \citet{shu88} in which a wind is driven from a boundary layer at
the interface between a star rotating near break--up velocity and an accretion disk. Many HAEBES
have measured \textit{v}sin\textit{i} values that are significant fractions of their break--up
velocities. These boundary layer outflows, however, also require a magnetic field at the interaction
region, although the required field strengths are small enough ($\sim$10--20 G for typical HAEBE
parameters; \citet{shu88}) to be within the current observational uncertainties for many objects
\citep{alecian13}. The high detection rate of PC profiles for accreting stars in our sample
indicates that stellar winds are common among accreting HAEBES.  Although our data are not
conclusive, the indication that the PC objects are not accreting magnetically lends itself to the
boundary layer accretion--driven stellar wind scenario.  These winds are likely generated partially
by the high rotation rates of HAEBES. Thus we tentatively conclude that boundary layer accretion in
HAEBES, assisted by centrifgual acceleration, is able to drive strong stellar winds.  

\subsection{HAEBES as massive CTTSs?}

HAEBES are often identified as the higher mass analogues of CTTSs
\citep[e.g.][]{muzz04,skinner04,grady10}. In reality, line profile statistics, including those
gathered here, point to significant differences between the incidence, and likely the physical
nature, of the mass flows in HAEBES compared to CTTSs. The optical permitted line study of CTTSs
performed by \citet{alencar00} found very high rates of outflow and accretion indicators, 80\% and
40\% respectively, in a sample of 30 CTTSs. A smaller sample of 15 objects was presented by
\citet{edwards94} with an even larger incidence (87\%, or 13 of 15 objects) of red--shifted
absorption in at least one line diagnostic. A large sample of HAEBES studied by \citet{fink84} found
only a 29\% incidence of blue--shifted absorption and only a 3\% incidence of red--shifted
absorption at H$\alpha$ or Na I. Blue--shifted forbidden line emission with both high and
low--velocity components is detected in almost all CTTSs \citep[e.g.][]{hartigan95}, indicating the
presence of extended collimated outflows. Similar line profiles are detected at a much lower rate in
HAEBES: \citet{cr97} found only 15\% of their 56 HAEBES display this characteristic while
\citet{bc94} find evidence of blue--shifted forbidden line emission in only 3\%of their sample. 

Our sample continues the trend of disparate line morphology statistics between HAEBES and CTTSs.  In
particular, outflows in both HAe and HBe stars seem to be much less common than in CTTSs while MA
signatures are present in a similar but slightly lower rate in HAe stars compared to CTTSs. HBe
stars show a considerably lower rate of red--shifted absorption signatures. These conclusions are
confirmed by the contingency test $p$--values given in \autoref{tab:tab6}. The tests provide strong evidence
that the rates of occurrence of these mass flow signatures differ between the two mass regimes and
perhaps between HAe and HBe stars, although the statistical evidence is weaker for the differences
between the HAEBE subgroups. Based on the \hei data then, it appears that the immediate
circumstellar environments of HAEBES and CTTSs cannot be treated identically as a general rule.  

The differences in \hei morphology statistics between HAEBES and CTTSs may be caused in part by
smaller magnetospheres mediating the star--disk interaction in HAe and some HBe stars, as suggested
by the results in \S 5.1.2. In addition, this would lead to a lower incidence of red--shifted absorption
signatures in HAEBES, which is observed, due to fewer viewing angles that intersect the infalling
part of the flow. The differing statistics may also be a result of the varying stages of pre--main
sequence evolution included in our sample: we simply may not be comparing objects at similar enough
stages of evolution, although we have attempted to remove the most obvious imposters from our
sample. More precise evolutionary subclasses for HAEBES, similar to those for WTTSs and CTTSs, may
account for the differences in line statistics observed here.

\begin{figure}\label{fig:fig9}
  \begin{center}
     \includegraphics[scale=.50,clip,trim=25mm 0mm 0mm 134mm]{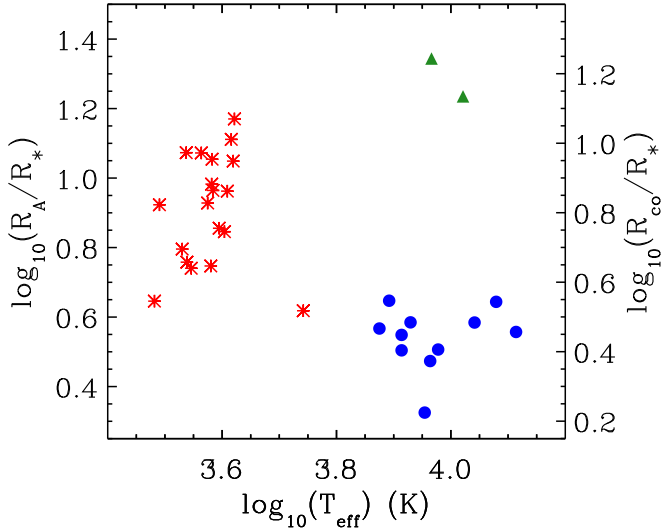}
     \caption{Alfv\'{e}n radius, $R_A$, plotted against $T_{eff}$ for the blue--shifted absorption
CTTSs from \citetalias{EFHK} (red asterisks) and the blue--shifted absorption HAEBES from this study
(blue circles and green triangles). The corotation radius, $R_{co}$, corresponding to each $R_A$ is
shown on the right--hand axis. The green triangles are V380 Ori and HD 190073, both confirmed
slowly rotating magnetic HAEBES. The HAEBES, in general, show smaller values of $R_A$ and $R_{co}$ 
which is consistent with the idea of smaller magnetospheres in HAEBES.}
  \end{center}
\end{figure}

As outlined in \S 2, smaller magnetospheres should result in smaller values of $R_A$. The ratio
$R_A$/$R_*$ is shown in \autoref{fig:fig9} for the CTTSs (red asterisks) exhibiting blue--shifted
absorption from \citetalias{EFHK} and for the HAEBES (blue circles and green triangles) from this
study showing blue--shifted absorption. The green triangles are V380 Ori and HD 190073 which are
both slowly rotating magnetic HAEBES. The right--hand axis shows the corotation radius. On average,
the HAEBES have smaller values of $R_A$, and thus smaller corotation radii, than the CTTSs. This is
mainly driven by the high $v$sin$i$ values of the HAEBES but this supports the idea that strong
outflows \textit{can} be launched from closer to the star and is consistent with the idea of smaller
magnetospheres on HAEBES. The \citet{shu88} X--celerator mechanism is, in fact, the limiting case of
a magnetosphere being pushed to the stellar surface with the outflow being launched centrifugally
along some of the open field lines. Thus it is not surprising that we see outflows in HAEBES that
are potentially launched from a point relatively closer to the stellar surface than seen in CTTSs. 

\section{SUMMARY AND FUTURE WORK}      

We have presented high resolution \hei observations of a large sample of Herbig Ae/Be stars.
These objects display a wide variety of line profile morphologies, including classical P--Cygni and
inverse P--Cygni profiles. Our results are summarized as follows: 

\begin{itemize}

\item A number of objects in our sample (7 of 56) do not display line profiles characteristic of
young stars interacting with their environments. These stars are flagged as having been potentially
misidentified as HAEBES. 

\item We find that there appear to be differences between the \hei line profile properties of HBe
stars compared to HAe stars. In particular, HBe objects tend to display blue absorption with little
evidence for infalling material; the HAe sample shows both blue and red shifted absorption features
indicating overall higher levels of mass flow activity. This may indicate that HBe objects do not
experience accretion in the same manner as HAe stars, i.e. instead of accreting along magnetic field
lines from a disk, accretion occurs through a boundary layer at the surface of the star.

\item HAe stars show a similar incidence of \hei red--shifted absorption to CTTSs while the
incidence in HBe stars is significantly lower. Both HAe and HBe stars show a much lower incidence of
blue--shifted absorption than is found in CTTSs. This suggests that HAe and especially HBe
environments cannot be treated as the intermediate mass analogs of CTTS systems. 

\item In objects that display red--shifted absorption, we observe no simultaneous blue--shifted
absorption, a feature common to CTTSs. Some blue absorption is expected (for most viewing angles)
due to winds that are launched from the inner disk or stellar surface. Since none of the objects in
our sample with clear red--shifted absorption show signs of blue absorption, it is unclear whether
or not these objects are capable of launching accretion--generated outflows.

\item The red--shifted absorption profiles in our sample show maximum absorption velocities that are
smaller percentages of the system escape velocity than is found in CTTSs, suggesting these mass
inflows originate closer to the star. This may be a result of more compact magnetospheres in HAEBES
resulting in smaller disk truncation radii. This is expected for the magnetospheric accretion
scenario for HAEBES due to their high rotation rates and weak or nonexistent magnetic fields.

\item None of the objects in our sample show any sign of narrow blue--shifted absorption indicative
of pure disk winds, which are common at \hei in CTTSs. We interpret these results as evidence that
the inner environments around HAEBES cannot be considered scaled analogues of CTTS environments.
More specifically, the magnetocentrifugally driven winds that are, theoretically, launched from near
the interaction region of the stellar magnetosphere and accretion disk in CTTSs are not clearly
indicated in HAEBES or are too weak to be detected in our data. This is supported by the lack of
detected magnetic fields on HAEBES. Instead, stellar winds seem to be the only outflow mechanism for
the objects in our sample. However, magnetocentrifugally driven ouflows from the star--disk
interaction region can appear as stellar winds if the interaction region is near enough to the star.
Thus these outflows cannot be ruled out based on the data presented here.  We find that for HAEBES
that do show outflow signatures, these can be driven from smaller magnetospheres than required for
CTTSs.  Finally, since stellar magnetic fields are either absent or too weak to be detected around
most HAEBES, and since we detect no disk wind signatures in our sample, we suggest that the magnetic
fields involved in launching disk winds in CTTSs are stellar in origin and not disk--generated or
primordial. This would naturally explain the lack of obvious disk winds in HAEBES and their common
appearance in CTTS systems.   

\end{itemize}

One potential cause for the differences between profile statistics in HAe and HBe stars and between
HAEBES and CTTSs is the likely spread of evolutionary stages present in the current catalog of
HAEBES. The more rapid evolution of HAEBES, and especially HBe stars, compared to CTTSs results in a
smaller chance of observing a particular HAEBE at any stage of its contraction towards the main
sequence. Thus it would be more informative to only compare HAe and HBe stars at the same stages of
evolution and to only compare CTTSs with HAEBES that are in the HAEBE equivalent of the CTTS phase.
\citet{malfait98} suggested an evolutionary classification for HAEBES based on IR colors similar to
the groupings given by \citet{hillenbrand92}. The relevant CTTS phase in the \citet{malfait98}
scheme would be panels (b) and (c) of their figure 3. A comparison of objects within these
sub--groups may be more precise than the analysis presented here.  However, a larger sample size
than the one presented in this work would be needed to make such a comparison meaningful.

As suggested by \citetalias{EFHK}, and numerically studied by \citet{kwan11}, simultaneous
measurements of He I $\lambda$5876 and \hei would provide constraints on the physical conditions of
\hei line formation. Data sets of this type, to our knowledge, have not been published. Similar
sample sizes of these data for CTTSs and HAEBES would facilitate an excellent comparison between the
wind launching and accretion regions for the two groups and help clarify the origin of the
statistical differences of line profile type presented in this work. We are presently pursuing this
for a small sample of HAEBES. 

The lack of observed wind signatures in the IPC objects in our sample is puzzling considering the
high rate (76\% from \citetalias{EFHK}) of CTTSs showing both red--shifted absorption and
blue-shifted absorption. We are currently analyzing a large sample of optical and UV HAEBE data. The
analysis presented here will be extended in these studies with the hope of shedding light on this
significant difference between HAEBES and CTTSs. In addition, variability studies of \hei in HAEBES
will help clarify whether simultaneous red and blue--shifted absorption is truly absent or if
our current data set is anomalous.     

\acknowledgements 

We wish to thank the staff of Kitt Peak National Observatory in Arizona for their hospitality and
help during the observing runs for this research. We also wish to thank the anonymous referee for
their careful reading and helpful comments that helped to improve this manuscript. A portion of our
analysis is based on observations obtained at the Gemini Observatory, which is operated by the
Association of Universities for Research in Astronomy, Inc., under a cooperative agreement with the
NSF on behalf of the Gemini partnership: the National Science Foundation (United States), the
National Research Council (Canada), CONICYT (Chile), the Australian Research Council (Australia),
Minist\'{e}rio da Ci\^{e}ncia, Tecnologia e Inova\c{c}\~{a}o (Brazil) and Ministerio de Ciencia,
Tecnolog\'{i}a e Innovaci\'{o}n Productiva (Argentina). This research has made use of the Simbad
Astronomical database and the NASA Astrophysics Data System. We wish to acknowledge partial support
for this research from the NSF through grant 1212122, from NASA Astrophysics Data Analysis Program
through grant NNX13AF09G, and from the Space Telescope Science Institute through grant
HST--60--12996.001 all made to Rice University. This study has made use of the SIMBAD database,
operated at CDS, Strasbourg, France and also NASA's Astrophysics Data System.

\end{document}